\documentclass{article}

\usepackage[utf8]{inputenc}
\usepackage[T1]{fontenc}
\usepackage{arxiv}
\usepackage{amsmath}
\usepackage{amsfonts}
\usepackage{amssymb}
\usepackage[authoryear,round]{natbib}
\usepackage{graphicx}
\usepackage[colorlinks,citecolor=blue,urlcolor=blue]{hyperref}
\usepackage{url}
\usepackage{booktabs}
\usepackage{nicefrac}
\usepackage{microtype}
\usepackage{doi}
\usepackage[figuresright]{rotating}
\usepackage{multirow}
\usepackage{caption}
\usepackage{subcaption}
\usepackage{longtable}
\usepackage{array}
\usepackage{makecell}
\usepackage{comment}
\usepackage[table]{xcolor}
\usepackage{soul}

\title{Average Cause-Specific Hazard: A Censoring-Invariant Measure of Event Burden Under Competing Risks}
\renewcommand{\shorttitle}{Average Cause-Specific Hazard}
\renewcommand{\headeright}{Nabi et al.}
\renewcommand{\undertitle}{}

\date{}

\author{
    {Khondoker~Nazmoon~Nabi} \\
    Department of Biostatistics \\
    Harvard T.H. Chan School of Public Health \\
    Boston, Massachusetts 02115, USA
    \And
    {Xiang~Meng} \\
    Department of Biostatistics \\
    Harvard T.H. Chan School of Public Health \\
    Boston, Massachusetts 02115, USA
    \And
    {Lu~Tian} \\
    Department of Biomedical Data Science \\
    Stanford University \\
    Stanford, California 94305, USA
    \And
    {Jean~M~Connors} \\
    Department of Hematologic Oncology \\
    Dana Farber Cancer Institute \\
    Boston, Massachusetts 02215, USA
    \And
    {Deb~Schrag} \\
    Department of Medicine \\
    Memorial Sloan Kettering Cancer Center \\
    New York, New York 10065, USA
    \And
    {Hajime~Uno}\thanks{Email: huno@ds.dfci.harvard.edu} \\
    Department of Data Science \\
    Dana Farber Cancer Institute \\
    Boston, Massachusetts 02215, USA
}

\usepackage{float}
\begin{document}
\maketitle

\begin{abstract}
\noindent Competing events are common in clinical and epidemiologic studies, including semi-competing risks in which a terminal event such as death may follow a nonfatal event but also competes with it beforehand. Standard summaries include the cumulative incidence function (CIF) and the incidence rate (IR), defined as the number of observed events divided by observed event-free person-time. With competing events, the naive IR generally depends on the censoring-time distribution unless intensities are constant. We propose the Average Cause-Specific Hazard (ACSH), a survival-weighted rate per event-free person-time that preserves the interpretation of an incidence rate and is defined purely from the event-time distribution, without involving the censoring-time distribution. We develop nonparametric estimation and inference for ACSH and, for two-sample comparisons, introduce ACSH differences and ratios that provide interpretable contrasts without requiring a strong model assumption between two groups. Simulation studies examine the finite-sample performance, and an analysis of the CANVAS trial illustrates the proposed methods.
\end{abstract}

\vspace{0.5cm}

%========================================================
\section{Introduction}
\label{sec:intro}
%========================================================
% 1) Competing risks are common; clinicians want interpretable effect magnitudes.
Competing events are ubiquitous in clinical and epidemiologic research. The occurrence of one event can preclude another or change how the latter should be interpreted. Competing risks may be classical, in which the possible event types are mutually exclusive, or semi-competing, in which a nonfatal event may occur before a terminal event but cannot occur afterward \citep{Pintilie2006-xf,FineJiangChappell2001}. Such data arise in oncology, cardiology, and geriatric research \citep{Kim2024,Lau2009}. The core challenge is to summarize event occurrence and treatment effects in ways that remain interpretable and well defined under censoring and competing events.

% 2) Quick review of the standard approach 
Competing risks data are typically summarized using cause-specific cumulative incidence functions (CIFs), which describe the probability of each event type over time \citep{AalenJohansen1978,Lin1997,Putter2007-un}. To summarize treatment effects, investigators often use the Fine–Gray model \citep{fine1999proportional}, reporting the subdistribution hazard ratio (sHR). Like the Cox proportional hazards model \citep{Cox1972}, the Fine–Gray model assumes that the ratio of subdistribution hazards is constant over time. When this assumption does not hold, the reported sHR can be difficult to interpret and may become sensitive to follow-up and censoring patterns \citep{Hernan2010}. A related issue, familiar from the Cox HR, is that a single sHR offers a convenient between-group summary but does not provide group-specific effect magnitudes on an absolute scale \citep{Uno2014moving}. Without accompanying group-specific risk or rate summaries, it is hard to assess whether the reported sHR reflects a clinically meaningful difference. 

% 3) About the incidence rate and the limitations of the traditional version
Another traditional approach summarizes event occurrence using incidence rates (IRs). The most common competing-risks implementation is a ``naive'' cause-specific IR: the number of observed events of a given cause divided by the total event-free person-time accumulated over follow-up. This quantity is easy to compute, but when cause-specific intensities vary over time, its population target depends on the censoring distribution even under independent censoring, as we illustrate in Section 3. As a result, the naive cause-specific IR and the between-group contrasts built from it may not correspond to clearly defined estimands.

% 4) Motivations
These considerations motivate population quantities that preserve the interpretability of a person-time rate and are defined solely from the event-time distribution. In single-event settings, the average hazard with survival weights (AH) has been proposed for this purpose \citep{Hajime2023,Uno2024-rg,Qian2025-bd,Horiguchi2026}. AH is a weighted average of the instantaneous intensity on $[0,\tau]$ with event-free-survival weights; as a population quantity, it is the event rate per unit event-free person-time and does not involve the censoring-time distribution. AH-based contrasts (differences and ratios) retain a direct rate-scale interpretation and, unlike proportional hazards summaries, do not require a proportionality assumption between groups. 

% 5) ACSH
In this paper we extend the AH framework to competing risks and introduce the Average Cause-Specific Hazard (ACSH). ACSH is a survival-weighted average of a cause-specific intensity per unit event-free person-time over a prespecified horizon $[0,\tau]$. Interpreted in words, ACSH is the average rate at which the event of interest occurs among subjects still free of any event, expressed per unit event-free person-time. As a population quantity, ACSH is defined entirely in terms of the event-time distribution; the censoring distribution does not enter its definition, while ACSH itself preserves the familiar interpretation of an IR. This distinguishes ACSH from the naive cause-specific IR: the latter is a sample quantity whose population limit involves the censoring-time distribution whenever the cause-specific intensities are non-constant and censoring is present. Corresponding two-sample contrasts of ACSH, the difference and the ratio, provide model-free rate-scale alternatives to the ratio of ``naive'' cause-specific IRs and the Fine–Gray sHR.

We retain from the AH framework the rate-scale definition, the influence-function representations that underlie variance estimation, and the delta-method construction of two-sample contrasts on natural and log scales. Our new contributions are (a)~ACSH, a new rate-scale summary metric for competing risks data; (b)~an extension to multiple non-terminal endpoints sharing a common terminal event; (c)~a scalar Total ACSH summary and its between-group contrasts; and (d)~a Wald-type global test across a prespecified set of endpoints or causes. Without competing events, ACSH reduces to the single-event AH of \citet{Hajime2023}.

\begin{figure}[ht]
\centering
\includegraphics[width=0.95\textwidth]{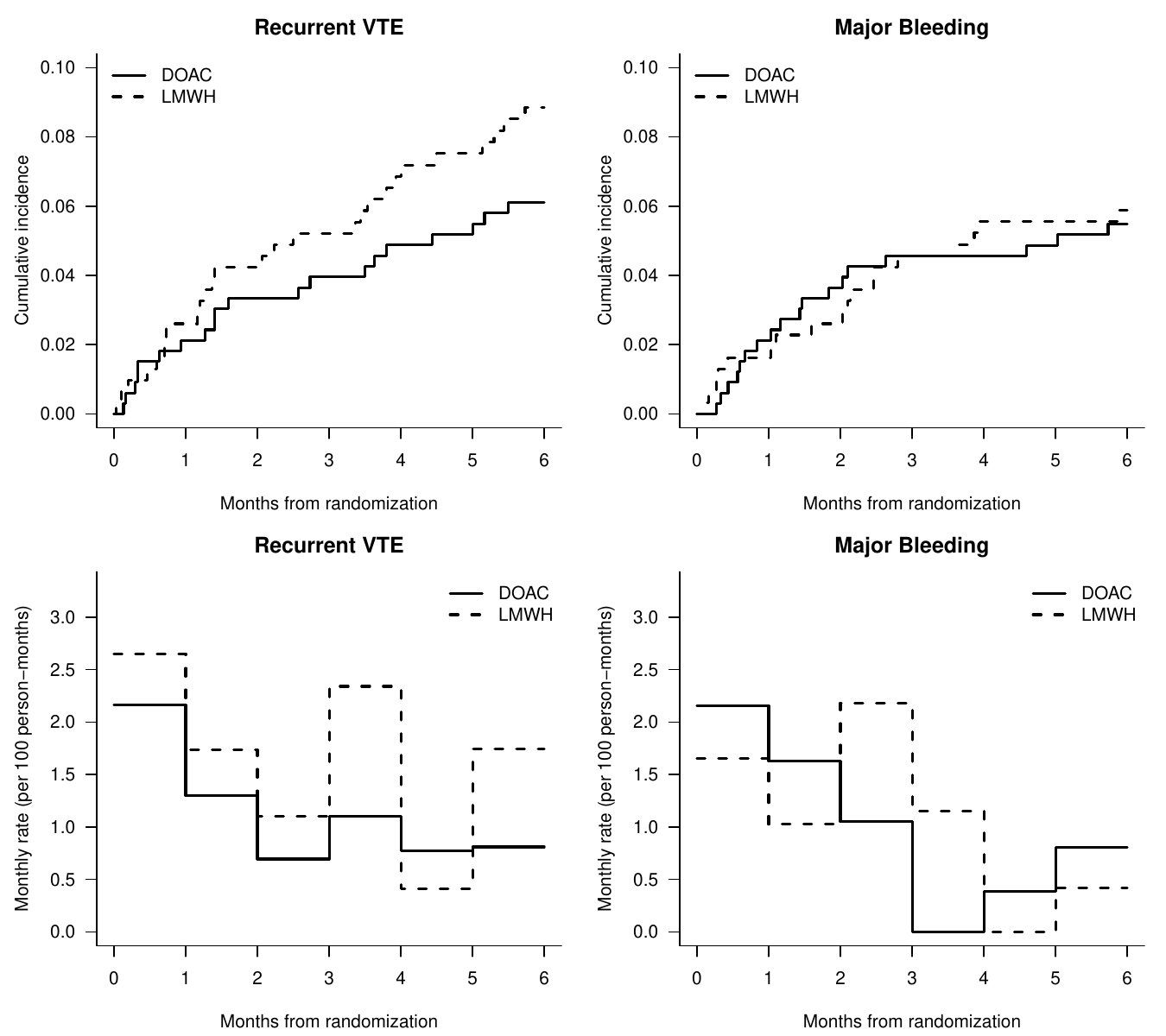}
\caption{CANVAS trial summaries over the 6-month follow-up period by treatment arm. \emph{Top row:} Aalen--Johansen cumulative incidence functions for recurrent VTE (left) and major bleeding (right), with death treated as the competing terminal event within each endpoint-specific analysis. \emph{Bottom row:} monthly piecewise-constant cause-specific rate for each endpoint, expressed per 100 person-months and computed for each month $m\in\{1,\ldots,6\}$ as the number of cause-specific events in $(m-1,m]$ divided by the total person-time accrued in the same interval; the rates decline over the follow-up horizon in both arms and for both endpoints, indicating time-varying cause-specific intensities. \emph{Abbreviations:} DOACs, direct oral anticoagulants; LMWH, low-molecular-weight heparin; VTE, venous thromboembolism.}
\label{fig:canvas_overview}
\end{figure}

%7) A brief introduction of the CANVAS trial as an illustrating example
To motivate and illustrate the proposed methods, we analyze data from CANVAS (Cancer-Associated Venous Thromboembolism Anticoagulation Strategies), a pragmatic randomized trial that compared direct oral anticoagulants (DOACs) with low-molecular-weight heparin (LMWH) among cancer patients with a newly diagnosed venous thromboembolism (VTE) \citep{Schrag2023-ym}. The trial followed 671 patients for up to 6 months and evaluated two clinically important non-terminal endpoints, recurrent VTE and major bleeding, with death as a common competing terminal event. Figure~\ref{fig:canvas_overview} (top row) displays the cumulative incidence functions for both endpoints by treatment arm, describing absolute risk over time. Monthly piecewise-constant hazard estimates (Figure~\ref{fig:canvas_overview}, bottom row) show that the cause-specific rates of both endpoints decline over the 6-month horizon, suggesting that the naive cause-specific IR does not target a censoring-invariant rate in this setting. We analyze CANVAS using ACSH-based one-sample summaries, two-sample contrasts, and a global Wald test across the two endpoints.

% 8) The paper structure  
The remainder of the paper proceeds as follows. Section~2 formalizes ACSH and develops its nonparametric estimation and inference. Section~3 reports simulation results. Section~4 presents the CANVAS analysis, and Section~5 gives concluding remarks.

%========================================================
\section{Methods}
\label{sec:methods}
%========================================================

\subsection{Setup and estimand}

Let $T\in[0,\infty)$ denote the time to the first event of any type, and let
$J\in\{1,\ldots,m\}$ denote the corresponding event type.  For cause $k$, let $\lambda_k(t)$ be the cause-specific hazard,
\[
\lambda_k(t)
=
\lim_{\epsilon\to 0}
\frac{\Pr\{t<T\le t+\epsilon,\;J=k\mid T\ge t\}}{\epsilon},
\]
and
$\Lambda_k(t)=\int_0^t \lambda_k(u)\,du$ be the corresponding cumulative cause-specific hazard.
Let
\(
S(t)=\Pr(T>t)
\)
denote the overall event-free survival function, that is, the probability of remaining free of any event up to time $t$. For a clinically meaningful horizon $\tau>0$, we also define the \emph{restricted mean event-free time} (RMEFT)
\begin{equation}
R(\tau)=\int_0^\tau S(u)\,du.
\label{eq:rmeft}
\end{equation}
This quantity represents the average time free of any event accumulated over $[0,\tau]$. The cumulative incidence function (CIF) for cause $k$ is
\begin{equation}
F_k(t)=\Pr(T\le t,\;J=k)=\int_0^t S(u)\,d\Lambda_k(u).
\label{eq:cif}
\end{equation}

With all preparations above, motivated by the average hazard with survival weights proposed by \citet{Hajime2023}, we define the
ACSH for cause $k$ over $[0,\tau]$ as
\begin{equation}
\eta_k(\tau)=\frac{F_k(\tau)}{R(\tau)}.
\label{eq:acsh_def}
\end{equation}
Thus, ACSH converts the probability-scale summary $F_k(\tau)$ into a rate-scale summary by dividing the cause-specific CIF by the average event-free time accumulated up to $\tau$. 
Equivalently, from (\ref{eq:cif}) and (\ref{eq:rmeft}), 
\[
\eta_k(\tau)=\frac{\int_0^\tau \lambda_k(u)S(u)\,du}{\int_0^\tau S(u)\,du}.
\]
As such, ACSH is a weighted average of the cause-specific hazard. In summary, ACSH is a restricted-time, event-free-probability-weighted rate summary.
Table~\ref{tab:acsh_simple} summarizes the key quantities used throughout.

%------------------------------------------
\begin{table}[t]
\centering
\caption{Summary measures for the Average Cause-Specific Hazard (ACSH) under competing risks.}
\label{tab:acsh_simple}
\small
\renewcommand{\arraystretch}{1.2}
\begin{tabular}{p{0.22\textwidth} p{0.31\textwidth} p{0.40\textwidth}}
\toprule
\textbf{Summary measure} & \textbf{Description} & \textbf{Interpretation} \\
\midrule
$F_k(\tau)$
& Cumulative Incidence Function (CIF) for cause $k$ evaluated at $\tau$
& Absolute risk: probability of experiencing a type-$k$ event by time $\tau$. \\

$R(\tau)=\int_0^\tau S(u)\,du$
& Restricted Mean Event-Free Time (RMEFT) up to $\tau$
& Average time free of any event accumulated over $[0,\tau]$. \\

$\eta_k(\tau)={F_k(\tau)}/{R(\tau)}$
& Average Cause-Specific Hazard (ACSH) for cause $k$ over $[0,\tau]$
& Average person-time incidence rate: the average number of type-$k$ events per event-free person-time over $[0,\tau]$. \\
\bottomrule
\end{tabular}
\end{table}
%------------------------------------------

\subsection{Nonparametric estimation and large-sample inference}
\label{sec:estimation}
In practice, the event time is oftentimes subject to right censoring. Let $C$ denote the right-censoring time.
We observe
\[
X=\min(T,C),\qquad K=J\,\mathbb{I}(T\le C),
\]
so that $K=0$ indicates right-censoring and $K=k\in\{1,\ldots,m\}$ indicates failure from cause $k,$
where $\mathbb{I}(A)$ denotes the indicator function, equal to 1 if $A$ is true and 0 otherwise.
We assume independent right-censoring, $C\perp (T,J),$ and 
$\Pr(C>\tau)>0$ for the prespecified truncation time $\tau>0.$  Observed data consist of $n$ independent and identically distributed (i.i.d.) copies of 
$(X, K):$ 
$$\left\{(X_i,K_i), i=1, \cdots, n \right\}.$$ 

For cause $k$, we estimate ACSH by
\[
\hat\eta_k(\tau)=\frac{\hat F_k(\tau)}{\hat R(\tau)}
\]
where
$$\hat F_k(\tau)=\int_0^\tau \hat S(u-)\,d\hat\Lambda_k(u), 
~~\mbox{  }~~\hat R(\tau)=\int_0^\tau \hat S(u)\,du,$$
$\hat S(t)$ is the Kaplan--Meier estimator of $S(t)$ obtained by treating \emph{any} event as a failure and
\[
\hat\Lambda_k(t)=\int_0^t \frac{d\bar N_k(u)}{\bar Y(u)}
\]
is the Nelson--Aalen estimator of the cause-specific cumulative hazard function
with
\[
\bar Y(t)=\frac{1}{n}\sum_{i=1}^n Y_i(t),
\qquad
\bar N_k(t)=\frac{1}{n}\sum_{i=1}^n N_{ki}(t),
\]
and
\(
Y_i(t)=\mathbb{I}(X_i\ge t)
\)
and
\(
N_{ki}(t)=\mathbb{I}(X_i\le t,\,K_i=k), k=1,\ldots,m,
\)
being the at-risk and counting processes for the $i$th observation, respectively.

Under the regularity conditions stated in Appendix~A, $\hat F_k(t)$ and $\hat{R}(t)$ are consistent estimators of $F_k(t)$ and $R(t)$, respectively, for $t\in [0,\tau].$ Consequently,
\[
\hat\eta_k(\tau)\to \eta_k(\tau)
\]
in probability, as $n \to \infty.$ Let
\[
\hat{\boldsymbol{\theta}}(\tau)
=
\bigl(\log\hat\eta_{1}(\tau),\ldots,\log\hat\eta_{m}(\tau)\bigr)^\top
\qquad \mbox{and} \qquad
\boldsymbol{\theta}(\tau)
=
\bigl(\log\eta_{1}(\tau),\ldots,\log\eta_{m}(\tau)\bigr)^\top.
\]
We have
\[
\sqrt{n}\{\hat{\boldsymbol{\theta}}(\tau)-\boldsymbol{\theta}(\tau)\}
=
\frac{1}{\sqrt{n}}\sum_{i=1}^n \boldsymbol{\psi}_i(\tau)+o_p(1)
\;\to\;
N_m\bigl(\mathbf{0},\boldsymbol{\Sigma}(\tau)\bigr),
\]
in distribution, as $n\to \infty,$ where $\boldsymbol{\psi}_i(\tau)$ is the influence function for the $i$th observation and
$\boldsymbol{\Sigma}(\tau)=\mathrm{Var}\{\boldsymbol{\psi}_i(\tau)\}$.  Specifically, $\boldsymbol{\psi}_i(\tau)=(\psi_{1,i}(\tau), \cdots, \psi_{m,i}(\tau))^\top,$ where 
\begin{equation}
\psi_{k,i}(\tau)
=
\int_0^\tau \frac{S(u)}{F_k(\tau)}\,\frac{dM_{ki}(u)}{G(u)}
+
\int_0^\tau \left\{\frac{F_k(u)}{F_k(\tau)}-\frac{R(u)}{R(\tau)}\right\}
\frac{dM_i(u)}{G(u)},
\label{eq:psi_logacsh_app}
\end{equation}
$M_{ki}(t)=N_{ki}(t)-\int_0^t Y_i(u)\,d\Lambda_k(u),$
$M_i(t)=\sum_{k=1}^m M_{k i}(t)$ and $G(t)=\Pr(X\ge t).$ Its derivation is given in Appendix~A.  Thus, a consistent estimator of $\boldsymbol{\Sigma}(\tau)$ is 
\[
\widehat{\boldsymbol{\Sigma}}(\tau)
=
\frac{1}{n}\sum_{i=1}^n \hat{\boldsymbol{\psi}}_i(\tau)^{\otimes 2},
\]
where $\hat{\boldsymbol{\psi}}_i(\tau)$ denotes the plug-in estimator of $\boldsymbol{\psi}_i(\tau)$, i.e., 
$\hat{\boldsymbol{\psi}}_i(\tau)=(\hat{\psi}_{1,i}(\tau), \cdots, \hat{\psi}_{m,i}(\tau))^\top,$  
\[
\hat{\psi}_{k,i}(\tau)
=
\int_0^\tau \frac{\hat{S}(u)}{\hat{F}_k(\tau)}\,\frac{d\hat{M}_{ki}(u)}{\hat{G}(u)}
+
\int_0^\tau \left\{\frac{\hat{F}_k(u)}{\hat{F}_k(\tau)}-\frac{\hat{R}(u)}{\hat{R}(\tau)}\right\}
\frac{d\hat{M}_i(u)}{\hat{G}(u)},
\]
$\hat{M}_{ki}(t)=N_{ki}(t)-\int_0^tY_i(u)d\hat{\Lambda}_k(u),$
$\hat{M}_i(t)=\sum_{k=1}^m \hat{M}_{k i}(t)$ and $\hat{G}(u)=\bar{Y}(u).$ Let $\widehat{\sigma}^2_k(\tau)$ denote the diagonal entry of $\widehat{\boldsymbol{\Sigma}}(\tau)$ corresponding to cause $k.$ The asymptotic variance of $\log\hat\eta_k(\tau)$ can be estimated by $\widehat{\sigma}^2_k(\tau)/n,$
and the corresponding $100(1-\alpha)\%$ confidence interval of $\eta_k(\tau)$ is
\[
\left[
\hat\eta_k(\tau)e^{-z_{1-\alpha/2}\frac{\widehat{\sigma}_k(\tau)}{\sqrt{n}}},
\;
\hat\eta_k(\tau)e^{z_{1-\alpha/2}\frac{\widehat{\sigma}_k(\tau)}{\sqrt{n}}}
\right].
\]

\paragraph{Remark 1.}
Noticing that $M_{ki}(t), k=1, \cdots, m$ are mutually orthogonal martingales, for $k, k'\in\{1,\ldots,m\}$ the $(k, k')$th element $\sigma_{kk'}(\tau)$ of $\boldsymbol{\Sigma}(\tau)$ can be further simplified to
\begin{equation}\sum_{\ell=1}^m \int_0^\tau \left[\xi_{k, \ell}(u, \tau)-\frac{R(u)}{R(\tau)}\right]\left[\xi_{k', \ell}(u, \tau)-\frac{R(u)}{R(\tau)}\right]\frac{d\Lambda_\ell(u)}{G(u)}, \label{eq:var-cmp}
\end{equation}
where
$$ \xi_{k, \ell}(u, \tau)=\frac{S(u)\mathbb{I}(\ell=k)+F_k(u)}{F_k(\tau)}, \quad k=1,\ldots,m.$$
Thus, $\boldsymbol{\Sigma}(\tau)$ can also be estimated by the corresponding plug-in estimator.

\subsection{Two-group comparisons}
\label{sec:twogroup}

Let $g\in\{0,1\}$ index two independent groups (e.g., treatment and control) with sample sizes $n_0$ and $n_1$, respectively. As described in (\ref{eq:cif}) and (\ref{eq:acsh_def}), for group $g,$ we may define the ACSH $\eta_{k,g}(\tau)$ based on the cumulative cause-specific hazard function $\Lambda_{k,g}(\cdot)$ and the RMEFT $R_g(\tau).$ 
We assume that observed data in group $g$ consists of $n_g$ i.i.d observations $(X_{g,i}, K_{g,i})_{i=1}^{n_g}$ and observations in group $g=0$ are independent of those in group $g=1$.
Within each group, we can obtain $$\hat{\boldsymbol{\theta}}_g(\tau)=(\log\hat{\eta}_{1,g}(\tau), \cdots, \log\hat{\eta}_{m,g}(\tau))^\top$$ 
to estimate 
$$\boldsymbol{\theta}_g(\tau)=(\log\eta_{1,g}(\tau), \cdots, \log\eta_{m,g}(\tau))^\top$$ 
as in section 2.2 and the variance of $\hat{\boldsymbol{\theta}}_g(\tau)$ can be estimated by 
\[
\widehat{\boldsymbol{\Sigma}}_g(\tau)
=
\frac{1}{n_g}\sum_{i=1}^{n_g}
\hat{\boldsymbol{\psi}}_{g,i}(\tau)\hat{\boldsymbol{\psi}}_{g,i}(\tau)^\top,
\]
where $\hat{\boldsymbol{\psi}}_{g,i}(\tau)$ is the plug-in estimate of the subject-level influence function in group $g$.

\paragraph{ACSH differences and ratios for a single cause}
For cause $k$, the between-group ACSH difference 
$$ D_k(\tau)=\eta_{k,1}(\tau)-\eta_{k,0}(\tau)$$
and 
ACSH ratio 
$$ Q_k(\tau)=\eta_{k,1}(\tau)/\eta_{k,0}(\tau) $$
can be consistently estimated by 
\[
\hat D_k(\tau)=\hat\eta_{k,1}(\tau)-\hat\eta_{k,0}(\tau)
~\mbox{ and }~
\hat Q_k(\tau)=\frac{\hat\eta_{k,1}(\tau)}{\hat\eta_{k,0}(\tau)},
\]
respectively. Furthermore, the variance of estimated between-group difference, $\hat{D}_k(\tau),$ can be estimated by 
\[
\sum_{g=0}^1 n_g^{-1}\hat\eta_{k,g}(\tau)^2
\widehat{\sigma}_{k,g}^2(\tau),
\]
where $\widehat{\sigma}_{k,g}^2(\tau)$ is the $k$th diagonal element of $\widehat{\boldsymbol{\Sigma}}_g(\tau).$
A Wald-type $100(1-\alpha)\%$ confidence interval for $D_k(\tau)$ is thus
\[
\left\{\hat D_k(\tau)-
z_{1-\alpha/2}\sqrt{\sum_{g=0}^1 \frac{\hat\eta_{k,g}(\tau)^2
\widehat{\sigma}_{k,g}^2(\tau)}{n_g}}, 
\hat D_k(\tau)+
z_{1-\alpha/2}\sqrt{\sum_{g=0}^1 \frac{\hat\eta_{k,g}(\tau)^2
\widehat{\sigma}_{k,g}^2(\tau)}{n_g}}\right\}.
\]
For the ACSH ratio, the inference is conducted on the log scale. Specifically, $\log Q_k(\tau)$ can be consistently estimated by $\log \hat{Q}_k(\tau),$ whose variance can be estimated by 
\[
\sum_{g=0}^1 n_g^{-1}\widehat{\sigma}_{k,g}^2(\tau).
\]
A $100(1-\alpha)$\% confidence interval for $Q_k(\tau)$ is 
$$
\left[\hat{Q}_k(\tau)e^{-z_{1-\alpha/2}\sqrt{\sum_{g=0}^1\frac{\widehat{\sigma}_{k,g}^2(\tau)}{n_g}}}, \hat{Q}_k(\tau)e^{z_{1-\alpha/2}\sqrt{\sum_{g=0}^1\frac{\widehat{\sigma}_{k,g}^2(\tau)}{n_g}}}\right].
$$
Both $\widehat{D}_k(\tau)$ and $\widehat{Q}_k(\tau)$ can be used to test  $H_0: \eta_{k,1}(\tau)=\eta_{k,0}(\tau)$. Specifically, a Wald test can be based on either 
$$Z_D=\frac{\widehat{D}_k(\tau)}{\sqrt{\sum_{g=0}^1 n_g^{-1}\hat{\eta}_{k,g}(\tau)^2\hat{\sigma}_{k,g}^2(\tau)}}$$
or
$$Z_Q=\frac{\log\{\widehat{Q}_k(\tau)\}}{\sqrt{\sum_{g=0}^1 n_g^{-1}\hat{\sigma}_{k,g}^2(\tau)}}.$$
The two tests are asymptotically equivalent.  
The null hypothesis is rejected at the two-sided significance level of $\alpha$, if the $Z$-score is greater than $z_{1-\alpha/2}$ or less than $z_{\alpha/2}.$

\paragraph{ACSH differences and ratios for multiple causes}
It is possible to simultaneously compare ACSH for multiple causes. A natural test statistic for global comparison is 
\[
\widehat{\boldsymbol{\Delta}}(\tau)
=
\hat{\boldsymbol{\theta}}_1(\tau)-\hat{\boldsymbol{\theta}}_0(\tau),
\]
whose variance can be estimated by 
\[
\widehat{\mathbf{V}}(\tau)
=
\frac{\widehat{\boldsymbol{\Sigma}}_1(\tau)}{n_1}
+
\frac{\widehat{\boldsymbol{\Sigma}}_0(\tau)}{n_0}.
\]
Under the null hypothesis
$H_0:\boldsymbol{\theta}_1(\tau)=\boldsymbol{\theta}_0(\tau)$,
\[
T_{\mathrm{global}}
=
\widehat{\boldsymbol{\Delta}}(\tau)^\top
\widehat{\mathbf{V}}(\tau)^{-1}
\widehat{\boldsymbol{\Delta}}(\tau)
\;\to\;
\chi^2_m
\]
in distribution, as $n \to \infty$. 
This provides an omnibus test of whether any log-ACSH differs between two groups. A similar method can be used to compare ACSH for a selected subset of causes by focusing on the corresponding sub-vector of $\hat{\boldsymbol{\Delta}}(\tau).$

\subsection{Extension: multiple non-terminal endpoints with a terminal event}
\label{sec:extension_multinonfatal}

In some applications, an individual may experience multiple clinically distinct non-terminal endpoints during follow-up,
whereas a terminal event such as death precludes any subsequent non-terminal event.
Recurrent VTE and major bleeding in the CANVAS trial \citep{Schrag2023-ym} are examples of non-terminal events.
Similarly, in cardiovascular studies, investigators often consider several nonfatal endpoints, such as heart failure,
hospitalization, and heart transplantation.
Because these non-terminal endpoints are not mutually exclusive, they cannot be fully represented as a single competing-risks outcome
$(T, J)$ as in the preceding sections.

Nevertheless, the ACSH framework extends naturally by analyzing each non-terminal endpoint through an endpoint-specific
time-to-first-event construction and treating death as the only competing terminal event for that endpoint.
Fix a non-terminal endpoint, e.g., VTE or major bleeding, indexed by $\ell \in\{1,\ldots, \tilde{m}\}$.
Let $T_\ell$ denote the time to event $\ell$, $D$ denote the time to death, and $C$ denote the censoring time. The observed data can be organized as $\{X_{1, i}, K_{1, i}, \cdots, X_{\tilde{m}, i}, K_{\tilde{m}, i}\}_{i=1}^n$: $n$ i.i.d. copies of $\{X_{1}, K_{1}, \cdots, X_{\tilde{m}}, K_{\tilde{m}}\},$ where
\[
X_\ell=\min(T_\ell,D,C) ~\mbox{and}~
K_\ell=
\begin{cases}
1, & T_\ell\le \min(D,C),\\
2, & D<\min(T_\ell,C),\\
0, & C<\min(T_\ell,D).
\end{cases}
\]
Importantly, occurrences of other non-terminal endpoints do \emph{not} remove a subject from the risk set for time to event $\ell$;
they are neither competing nor censoring events for the endpoint-$\ell$ analysis. For event $\ell$, define
\[
F_{1,\ell}(t)=\Pr(T_\ell\le t,\;T_\ell<D),
\qquad
S_\ell(t)=\Pr(T_\ell\wedge D>t)
\qquad \mbox{and} \qquad
R_\ell(\tau)=\int_0^\tau S_\ell(u)\,du,
\]
which is the restricted mean endpoint-$\ell$-free time. Then, the event $\ell$-specific ACSH is
\begin{equation}
\eta_\ell(\tau)=\frac{F_{1,\ell}(\tau)}{R_\ell(\tau)},
\label{eq:acsh_multinonfatal}
\end{equation}
which is the average rate of first occurrence of endpoint $\ell$ per unit time spent alive and event $\ell$ free over the time window $[0,\tau]$.

For each event $\ell$, statistical inference on $\eta_\ell(\tau)$ proceed exactly as in Sections~\ref{sec:estimation} and~\ref{sec:twogroup},
treating $(X_\ell,K_\ell)$ as a two-cause competing-risks outcome:
cause 1 = endpoint $\ell$, cause 2 = death before endpoint $\ell$, with right-censoring when $K_\ell=0$.
Accordingly, one can obtain a consistent estimator of $\eta_\ell(\tau)$, $\hat\eta_\ell(\tau)$ based on $\{X_{\ell, i}, K_{\ell, i}\}_{i=1}^n.$  Let 
\[
\hat{\boldsymbol{\theta}}^{\mathrm{NT}}(\tau)
=
\bigl(\log\hat\eta_1(\tau),\ldots,\log\hat\eta_{\tilde{m}}(\tau)\bigr)^\top
\]
and
\[
{\boldsymbol{\theta}}^{\mathrm{NT}}(\tau)
=
\bigl(\log\eta_1(\tau),\ldots,\log\eta_{\tilde{m}}(\tau)\bigr)^\top.
\]
As in Sections~\ref{sec:estimation} and~\ref{sec:twogroup}, we also have
\[
\sqrt{n}\{\hat{\boldsymbol{\theta}}^{\mathrm{NT}}(\tau)-\boldsymbol{\theta}^{\mathrm{NT}}(\tau)\}=\frac{1}{\sqrt{n}}\sum_{i=1}^n \hat{\boldsymbol{\psi}}^{\mathrm{NT}}_i(\tau)+o_p(1),
\]
where $\boldsymbol{\psi}_i^{\mathrm{NT}}(\tau)=(\psi_{1,i}^{\mathrm{NT}}(\tau), \cdots, \psi_{\tilde m,i}^{\mathrm{NT}}(\tau))^\top,$ 
\[
\psi_{\ell,i}^{\mathrm{NT}}(\tau)
=
\int_0^\tau \frac{S_\ell(u)}{F_{1,\ell}(\tau)}\,\frac{dM_{1, \ell, i}(u)}{G_\ell(u)}
+
\int_0^\tau \left\{\frac{F_{1,\ell}(u)}{F_{1, \ell}(\tau)}- \frac{R_\ell(u)}{R_{\ell}(\tau)}\right\}
\frac{dM_{\ell, i}(u)}{G_\ell(u)},
\]
\[
M_{1, \ell, i}(t)=\mathbb{I}(X_{\ell, i}\le t, K_{\ell, i}=1)-\int_0^{t\wedge X_{\ell, i}} \lambda_{1,\ell}(u)du,
\]
\[
M_{\ell, i}(t)=\mathbb{I}(X_{\ell, i}\le t, K_{\ell, i}>0)-\int_0^{t\wedge X_{\ell, i}}\lambda_\ell(u)du,
\]
\[
\lambda_{1, \ell}(t)=\lim_{\epsilon \downarrow 0} \frac{P(T_\ell\wedge D<t+\epsilon, T_\ell<D \mid T_\ell\wedge D\ge t)}{\epsilon}
\]
\[
\lambda_\ell(t)=\lim_{\epsilon \downarrow 0} \frac{P(T_\ell\wedge D<t+\epsilon\mid T_\ell\wedge D\ge t)}{\epsilon}
\]
and
$G_\ell(t)=\Pr(X_\ell\ge t).$ 
Therefore, as $n \to \infty,$ $\sqrt{n}\{\hat{\boldsymbol{\theta}}^{\mathrm{NT}}(\tau)-\boldsymbol{\theta}^{\mathrm{NT}}(\tau)\}$ converges in distribution to a multivariate Gaussian with mean zero and a variance-covariance matrix of $\boldsymbol{\Sigma}^{\mathrm{NT}}(\tau),$ which can be consistently estimated by 
\[
\widehat{\boldsymbol
{\Sigma}}^{\mathrm{NT}}(\tau)
=
\frac{1}{n}\sum_{i=1}^n
\hat{\boldsymbol{\psi}}^{\mathrm{NT}}_i(\tau)
\hat{\boldsymbol{\psi}}^{\mathrm{NT}}_i(\tau)^\top,
\]
where
$\hat{\boldsymbol{\psi}}^{\mathrm{NT}}_i(\tau)
=
\bigl(\hat\psi_{1,i}^{\mathrm{NT}}(\tau),\ldots,\hat\psi_{\tilde{m},i}^{\mathrm{NT}}(\tau)\bigr)^\top
$
and $\hat\psi_{\ell,i}^{\mathrm{NT}}(\tau)$ is a plug-in estimator of $\psi_{\ell, i}^{\mathrm{NT}}(\tau)$ by replacing all unknown quantities by their respective estimators. Note that 
$\boldsymbol{\Sigma}^{\mathrm{NT}}(\tau)$ does not have a similar simplification in (\ref{eq:var-cmp}) as the martingale central limit theorem is not applicable in this case.

\paragraph{Total ACSH summary and between-group comparisons}
With established large sample properties of $\hat{\boldsymbol{\theta}}^{\mathrm{NT}}(\tau)$, an omnibus test for two-group comparison across multiple non-terminal events can then be conducted similarly as in Section~\ref{sec:twogroup}. On the other hand, it is desirable to summarize the total disease burden due to a set of non-terminal events of interest and evaluate the treatment effect accordingly.  To this end, let
\begin{equation}
\eta_{\mathrm{tot}}(\tau)=\sum_{\ell=1}^{\tilde{m}}\eta_\ell(\tau)
=
\mathbf{1}^\top \boldsymbol{\eta}^{\mathrm{NT}}(\tau),
\label{eq:acsh_total}
\end{equation}
where $\mathbf{1}$ denotes the $\tilde{m}$-dimensional vector of ones.
Because each component ACSH uses its own endpoint-specific denominator $R_\ell(\tau)$, $\eta_{\mathrm{tot}}(\tau)$ is best viewed
as a scalar summary on the ACSH scale rather than as a single common-denominator incidence rate.  $\eta_{\mathrm{tot}}(\tau)$ summarizes multi-endpoint burden by combining endpoint-specific first-occurrence rates on the ACSH scale. This interpretation is more appropriate than the summary in Section~\ref{sec:twogroup} when the non-terminal endpoints are not mutually exclusive, because Section~\ref{sec:twogroup} only yields a summary for the first-composite event. Although these two summaries may coincide in simple settings such as independent exponential event times, they generally answer different scientific questions; see Appendix A for a simple illustration.

To make statistical inference on $\eta_{\mathrm{tot}}(\tau)$, let
\(
\hat{\boldsymbol{\eta}}^{\mathrm{NT}}(\tau)
=
\bigl(\hat\eta_1(\tau),\ldots,\hat\eta_{L^\ast}(\tau)\bigr)^\top.
\)
By the multivariate delta method, the variance of $\hat\eta_{\mathrm{tot}}(\tau)$ is approximately 
\[
n^{-1}\,
{\boldsymbol{\eta}}^{\mathrm{NT}}(\tau)^\top
\boldsymbol{\Sigma}^{\mathrm{NT}}(\tau)
{\boldsymbol{\eta}}^{\mathrm{NT}}(\tau),
\]
which can be estimated by 
\[
n^{-1}\,
\hat{\boldsymbol{\eta}}^{\mathrm{NT}}(\tau)^\top
\widehat{\boldsymbol{\Sigma}}^{\mathrm{NT}}(\tau)
\hat{\boldsymbol{\eta}}^{\mathrm{NT}}(\tau).
\]
A Wald-type confidence interval for $\eta_{\mathrm{tot}}(\tau)$ follows immediately.

For two-group comparison, define $\hat\eta_{\mathrm{tot},g}(\tau)$ analogously within group $g\in \{0, 1\}.$  
The between-group Total ACSH difference is
\[
D_{\mathrm{tot}}(\tau)=\eta_{\mathrm{tot},1}(\tau)-\eta_{\mathrm{tot},0}(\tau),
\]
which can be consistently estimated by
\[
\hat D_{\mathrm{tot}}(\tau)=\hat\eta_{\mathrm{tot},1}(\tau)-\hat\eta_{\mathrm{tot},0}(\tau).
\]
By independence of the two groups and the delta method,
\[
\hat D_{\mathrm{tot}}(\tau)-D_{\mathrm{tot}}(\tau)
\]
is approximately mean-zero normal with variance estimated by
\[
\widehat{\sigma}^2_{\mathrm{tot}}(\tau)
=
\sum_{g=0}^1
n_g^{-1}
\hat{\boldsymbol{\eta}}^{\mathrm{NT}}_g(\tau)^\top
\widehat{\boldsymbol{\Sigma}}^{\mathrm{NT}}_g(\tau)
\hat{\boldsymbol{\eta}}^{\mathrm{NT}}_g(\tau).
\]
A Wald-type confidence interval for $D_{\mathrm{tot}}(\tau)$ and the corresponding Wald test of $H_0:\eta_{\mathrm{tot},1}(\tau)=\eta_{\mathrm{tot},0}(\tau)$, follow immediately. A ratio of total ACSH can be handled analogously.

%===============================================
\section{Simulation Study}\label{sec:simulation}
We conducted a simulation study to evaluate the finite-sample performance of the proposed ACSH estimator and to illustrate its advantage over the naive incidence rate (IR) in competing risks settings. Specifically, we examined operating characteristics under both constant-hazard and time-varying-hazard settings, focusing on bias, precision, standard error calibration, and confidence interval coverage. The non-exponential scenario was included to demonstrate that, when hazards vary over time, the naive IR's population target depends on the censoring distribution, whereas ACSH remains well defined and interpretable on the event-free person-time scale.

\subsection{Setup}\label{sec:simulation:setup}
We considered a semi-competing risks model with two non-fatal events of interest (Causes~1 and~2) and a terminal event, death, under independent right-censoring and a prespecified truncation horizon $\tau=5,$ 
For each simulated dataset, we estimated the cause-specific ACSH for causes 1 and 2 using the proposed inference procedure. As a comparator, we also computed the naive cause-specific incidence rate, defined as the number of observed events of a given cause divided by the total observed event-free person-time accumulated up to $\tau$, i.e. 
$$ \widehat{\eta}_{\ell}^{\mathrm{naive}}(\tau)=\frac{\sum_{i=1}^n \mathbb{I}(X_{\ell, i}\le \tau, K_{\ell, i}=1)}{\sum_{i=1}^n \min(X_{\ell, i}, \tau)}, \ell=1, 2.$$

The performance of estimation methods was evaluated in terms of relative bias, root mean squared error (RMSE), average estimated standard error (ASE), empirical standard error (ESE), and empirical coverage of nominal 95\% Wald confidence intervals.

We examined three data-generating scenarios. In Scenario~(i) two independent latent event times and time to a terminal event (death) followed independent exponential distributions,
\[
T_1 \sim \mathrm{Exp}(\lambda_1), ~~ T_2 \sim \mathrm{Exp}(\lambda_2) ~~\mbox{and}~~ D \sim \mathrm{Exp}(\lambda_D)
\]
with $\lambda_1=0.15,$ $\lambda_2=0.25$ and $\lambda_D=0.20$; and censoring times followed $C \sim \mathrm{Exp}(\mu)$ with $\mu=0.10$. Consistent with the endpoint-specific construction of Section~\ref{sec:extension_multinonfatal}, each event of interest $\ell\in\{1,2\}$ is observed as its own two-cause competing-risks outcome with death: $X_\ell=\min(T_\ell, D, C)$, and $K_\ell=1,$ if $T_\ell\le\min(D,C)$ (event $\ell$), $K_\ell=2,$ if $D<\min(T_\ell,C)$ (death), and $K_\ell=0,$ otherwise (censored). Occurrences of the other event of interest do not remove a subject from the risk sets for endpoint $\ell$.
The true ACSH value for Causes~$\ell$ was $\lambda_\ell$, because
$$F_{1, \ell}(\tau)=\int_0^\tau \lambda_\ell e^{-(\lambda_{\ell}+\lambda_D)t}dt, R_{\ell}(\tau)=\int_0^\tau e^{-(\lambda_{\ell}+\lambda_D)t}dt~~\mbox{and}~~  \eta_{\ell}(\tau)=\frac{F_{1, \ell}(\tau)}{R_\ell(\tau)}=\lambda_\ell.$$ 
This simple setting serves as a benchmark, where the naive estimator is expected to perform well.

In Scenarios~(ii) and (iii), we introduced time-varying hazards by generating latent event times from Gamma distributions with a common shape parameter of 1.5 and scale parameters of 4.44 and 2.67 for Causes~1 and 2, respectively, with the terminal event (death) following $D \sim \mathrm{Exp}(\lambda_D)$, where $\lambda_D=0.10.$
The true ACSH value is 0.1195 for Cause~1 and 0.2206 for Cause~2 based on numerical integration. In Scenario~(ii), all observations are followed up to $\tau$ without additional censoring. In Scenario~(iii), there is additional censoring induced by an independent $C \sim \mathrm{Exp}(\mu)$ with $\mu=0.10$. The comparison between Scenarios~(ii) and (iii) isolates the role of random censoring and shows that the naive estimator becomes distorted in the presence of both time-varying hazards and random censoring.

For all three scenarios, we considered sample sizes $n \in \{300,1000\}$ and used $R=1,000$ Monte Carlo replicates per configuration. To keep the presentation compact, Table~\ref{tab:sim_metrics} reports results for Cause~1 only; the full results, including Cause~2, are provided in 
Appendix~B.

\subsection{Results}\label{sec:simulation:results}

Across all three scenarios, the proposed ACSH estimator performed well. Relative bias was negligible, the RMSE decreased at the expected rate as $n$ increased from 300 to 1,000, ASE and ESE were closely aligned, indicating a good accuracy of the influence function-based variance estimator, and the empirical coverage of nominal 95\% Wald confidence intervals remained close to the nominal level throughout (Table~\ref{tab:sim_metrics}).

The naive estimator behaved quite differently across the three scenarios. Under Scenario~(i), with constant hazards and random censoring, the naive estimator targets the true ACSH: relative bias was negligible and coverage level of the corresponding confidence intervals remained close to 0.95. Under Scenario~(ii), with time-varying hazards but no random censoring, the naive estimator and the proposed estimator are identical. The informative comparison arises in Scenario~(iii), where time-varying hazards and random censoring present simultaneously. In this setting, the naive estimator exhibited a persistent negative relative bias of approximately $-3.8\%$ to $-4.0\%$ that did not diminish with increasing sample size, while the empirical coverage of its 95\% confidence interval decreased to 0.900 at $n=1000$. In contrast, the ACSH estimator remained essentially unbiased.

These observations are consistent with the fact that, under independent random censoring, the estimand of the naive estimator is
\[
\tilde{\eta}_k(\tau)
=
\frac{\int_0^\tau G_C(u)\,dF_k(u)}
{\int_0^\tau S(u)\,G_C(u)\,du}
=
\frac{\int_0^\tau S(u)G_C(u)\lambda_k(u)\,du}
{\int_0^\tau S(u)G_C(u)\,du},
\]
where $G_C(\cdot)$ denotes the survival function of the censoring time. This ratio coincides with $\eta_k(\tau)$ when either the cause-specific hazard $\lambda_k(u)$ is constant over $[0,\tau]$ (Scenario~(i)) or $G_C(u)\equiv 1$ (Scenario~(ii)), but differs from $\eta_k(\tau)$, otherwise, consistent with the observed bias in Scenario~(iii) \citep{Hajime2023}.

In summary, the bias of the naive estimator in Scenario~(iii) arises from the difference between the weighting functions $S(u)G_C(u)$ and $S(u)$. In the practically most relevant setting, where hazards vary over time and the data are subject to right censoring, the naive estimator becomes systematically biased and its confidence intervals exhibit undercoverage for the true ACSH. In contrast, the proposed ACSH estimator remains approximately unbiased with valid confidence interval, providing a robust approach for studying this censoring-invariant summary of cause-specific event occurrence per unit event-free person-time over a fixed follow-up horizon.
The robustness of these conclusions was further examined in Appendix~C. The confidence intervals for ACSH retain a near nominal coverage level under increasing and decreasing hazards, censoring fractions varying from 12\% to 66\%, and cause-specific cumulative incidence rates as low as 3\%, whereas the inference based on the naive estimator may fail.

\subsection{Additional simulation results for two-sample contrasts}

The simulations in Sections~\ref{sec:simulation:setup}--\ref{sec:simulation:results} focused on one-sample performance of the proposed ACSH estimator and on its contrast with the naive IR. To complement these results, additional simulation results are provided in Appendix~D. These supplementary results were obtained using a simulation framework calibrated to the observed CANVAS data and report the finite-sample performance of the endpoint-specific arm-wise ACSH estimates, their between-group differences and ratios, and the analogous quantities for the Total ACSH summary introduced in Section~2.4.

Specifically, Appendix D reports performance measures for the arm-specific ACSH estimates in recurrent VTE and major bleeding and for the corresponding DOAC versus LMWH contrasts, together with the Total ACSH arm estimates and contrasts. These analyses were included to verify that the proposed large-sample inference procedure remains well calibrated not only for one-sample ACSH estimation but also for endpoint-specific two-sample comparisons and for the scalar Total ACSH summary in realistic sample sizes.

Because these additional results are primarily confirmatory and do not change the main message of the paper, we defer the numerical details to Appendix~D.
In addition, Appendix~E reports a simulation study with nonzero cross-endpoint correlation, in which the finite-sample validity of the estimators for covariance matrix $\boldsymbol\Sigma^{\mathrm{NT}}$ (including its off-diagonal entries) and the variance of Total ACSH is confirmed.

%==================================
\section{Example}
\label{sec:example}
\subsection{Data from the CANVAS trial}
As an illustrative application, we analyzed data from the CANVAS trial (Cancer-Associated Venous Thromboembolism Anticoagulation Strategies), a pragmatic, multicenter randomized clinical trial comparing direct oral anticoagulants (DOACs) with low-molecular-weight heparin (LMWH) in patients with cancer-associated venous thromboembolism (VTE) \citep{Schrag2023-ym}. The trial was conducted across 67 oncology practices in the United States between 2016 and 2020 and enrolled 671 adult patients with active cancer and a newly diagnosed VTE. Participants were randomized in a 1:1 ratio to receive either DOAC or LMWH and were followed for up to 6 months or until death.
Consistent with the primary (as-treated) analysis population of the CANVAS trial, our analysis was based on the 638 randomized participants who initiated their assigned treatment.

For the present analysis, we considered two clinically important non-terminal endpoints: recurrent VTE and major bleeding within 6 months. Death was treated as a terminal competing event for each endpoint-specific analysis. Thus, consistent with the framework in Section~\ref{sec:extension_multinonfatal}, recurrent VTE and major bleeding were not analyzed as mutually exclusive competing events with one another. Instead, each endpoint was analyzed separately using a time-to-first-event formulation, with death as the competing terminal event and loss to follow-up treated as censoring.

\subsection{Endpoint-specific and joint ACSH analyses}
We analyzed recurrent VTE and major bleeding as two distinct non-terminal endpoints, using the endpoint-specific framework described in Section~\ref{sec:extension_multinonfatal}. For each endpoint, time to first occurrence of that endpoint was analyzed with death treated as the competing terminal event. Thus, recurrent VTE and major bleeding were not treated as mutually exclusive competing events with one another; rather, each endpoint was evaluated separately, and joint inference was based on the covariance of the corresponding endpoint-specific ACSH estimators.

Figure~\ref{fig:canvas_overview} (top row) shows the Aalen--Johansen estimates of the endpoint-specific cumulative incidence functions for the DOAC and LMWH groups over the 6-month follow-up period. By 6 months, the estimated cumulative incidence of recurrent VTE was slightly lower in the DOAC group than in the LMWH group, while the cumulative incidence of major bleeding was similar between the two arms. The CIF curves describe absolute risk over time and provide a probability-scale context for interpreting the ACSH-based rate summaries that follow.

Table~\ref{tab:acsh_comparison_canvas} summarizes the endpoint-specific ACSH estimates for the DOAC and LMWH groups over the 6-month follow-up period. For recurrent VTE, the estimated ACSH was 1.17 per 100 person-months of event-free follow-up (95\% CI: 0.76, 1.82) in the DOAC group and 1.70 per 100 person-months (95\% CI: 1.16, 2.48) in the LMWH group, corresponding to an ACSH difference of $-0.52$ per 100 person-months (95\% CI: $-1.35$, 0.30) and an ACSH ratio of 0.692 (95\% CI: 0.388, 1.237). For major bleeding, the estimated ACSH was 1.06 per 100 person-months (95\% CI: 0.66, 1.69) in the DOAC group and 1.11 per 100 person-months (95\% CI: 0.70, 1.76) in the LMWH group, yielding an ACSH difference of $-0.05$ per 100 person-months (95\% CI: $-0.76$, 0.66) and an ACSH ratio of 0.955 (95\% CI: 0.495, 1.844). For both endpoints, the point estimates favored the DOAC group, although the confidence intervals included the null.

To assess treatment effects jointly across the two non-terminal endpoints, we applied the global Wald test described in Section~\ref{sec:twogroup}. The resulting test statistic was $T_{\mathrm{global}}=1.55$ on two degrees of freedom, corresponding to a $p$-value of 0.46. Thus, the data do not provide evidence of a joint between-group difference in the endpoint-specific ACSH values for recurrent VTE and major bleeding.

As a secondary scalar summary, we also computed Total ACSH, defined as the sum of the endpoint-specific ACSH values. The estimated Total ACSH was 2.23 per 100 person-months of event-free follow-up (95\% CI: 1.48, 2.98) in the DOAC group and 2.80 per 100 person-months (95\% CI: 1.91, 3.70) in the LMWH group, corresponding to a difference of $-0.57$ per 100 person-months (95\% CI: $-1.74$, 0.60) and a ratio of 0.796 (95\% CI: 0.501, 1.267). Because the component ACSHs use endpoint-specific denominators, Total ACSH is best interpreted as a scalar summary on the ACSH scale rather than as a single common-denominator incidence rate. For example, the LMWH Total ACSH of 2.80 per 100 person-months should not be read as 2.80 combined events per 100 person-months alive and free of \emph{all} endpoints; each component ACSH is defined on its own event-free person-time denominator, and Total ACSH is simply the sum of those component rates. In the present data, this summary was lower in the DOAC group, although the associated uncertainty remained substantial.

%=================================
\section{Remarks}

We proposed the ACSH, $\eta_k(\tau)$, as a nonparametric cause-specific rate-scale summary for competing risks data. As a population quantity, ACSH is defined solely from the event-time distribution and preserves the familiar interpretation of an incidence rate.

ACSH is intended to complement, rather than replace, the CIF. CIF remains the natural summary for absolute risk over time, whereas ACSH provides a summary on the rate scale. Reporting both can therefore be useful: the CIF describes how likely an event is to occur by time $\tau$, while ACSH describes how frequently it occurs per unit event-free time over the same horizon. In two-sample settings, ACSH differences and ratios provide interpretable contrasts without requiring proportional subdistribution hazards.

A practical advantage of ACSH is that it can be estimated using standard nonparametric methods for competing risks data. Specifically, estimation requires only the Aalen--Johansen estimator for the CIF and the Kaplan--Meier estimator for the event-free survival function as building blocks. The corresponding influence-function representation yields large-sample inference for one-sample ACSH, two-sample ACSH contrasts, and Wald tests across multiple event types.

The choice of the truncation time $\tau$ is important. As with other restricted-time summaries, ACSH should always be interpreted relative to the chosen horizon. In practice, $\tau$ should be selected based on the scientific question and the available follow-up, avoiding values so large that few subjects remain under observation.

The extension to multiple non-terminal endpoints broadens the scope of the proposed framework. In such settings, each endpoint can be analyzed through an endpoint-specific time-to-first-event construction with death as the competing terminal event. This yields endpoint-specific ACSH estimands and permits joint inference across multiple non-terminal events.

We also considered a Total ACSH obtained by summing endpoint-specific ACSH values. Because the component ACSHs generally involve different endpoint-specific denominators, this quantity is best viewed as a scalar summary on the ACSH scale rather than as a single common-denominator incidence rate. 
As reported in Appendix~D, confidence intervals for Total ACSH and for its between-group contrast attain empirical coverage close to the nominal level in all settings investigated, supporting its use in practice.

We emphasize that ACSH is built from the cause-specific hazard and is distinct from the subdistribution hazard. An analogous restricted-time rate-scale summary could be developed from the subdistribution hazard, with corresponding one-sample, two-sample, and regression-based inference procedures; its development is a natural direction for future work but is beyond the scope of this paper \citep{PutterSchumacher2020}.

Several limitations should be noted. Our development assumes independent right-censoring. If this assumption fails, the plug-in estimator is no longer consistent for $\eta_k(\tau)$, and extensions to relax this assumption (for example, via inverse probability of censoring weighting) would be required. In addition, although ACSH is an interpretable scalar summary, it does not describe the full temporal pattern of treatment effects; for that reason, it is best interpreted alongside graphical summaries such as CIF curves. Future work could consider regression modeling for ACSH and weighted summaries across multiple non-fatal endpoints.

% R package
The \texttt{survACSH} R package, which implements one-sample, two-sample, Total ACSH, and global Wald inference, will be made available to users upon request.

%=============================================

\newpage
\begin{table}[ht]
\centering
\caption{\textbf{Cause~1 only.} Monte Carlo operating characteristics of the proposed average cause-specific hazard (ACSH) estimator and the naive person-time incidence rate (Naive IR) under three data-generating scenarios: (i)~exponential latent event times with independent exponential censoring, (ii)~Gamma latent event times with no censoring, and (iii)~Gamma latent event times with independent exponential censoring. Latent event times for Causes~1 and~2 were $T_1\sim\mathrm{Exp}(0.15)$ and $T_2\sim\mathrm{Exp}(0.25)$ in Scenario~(i), and independent $\mathrm{Gamma}(1.5,\,4.44)$ and $\mathrm{Gamma}(1.5,\,2.67)$ in Scenarios~(ii)--(iii). When present, censoring followed $C\sim\mathrm{Exp}(0.10)$. 
A terminal competing event (death) followed $\mathrm{Exp}(0.20)$ in Scenario~(i) and $\mathrm{Exp}(0.10)$ in Scenarios~(ii)--(iii). 
Truncation time $\tau=5$, sample sizes $N\in\{300,1000\}$, and $R=1000$ Monte Carlo replicates per configuration. True ACSH for Cause~1 is 0.15 in Scenario~(i) and 0.1195 in Scenarios~(ii)--(iii). Relative Bias denotes the Monte Carlo mean of $(\hat{\theta}-\theta)/\theta$; RMSE is the root mean squared error; ASE is the average estimated standard error; ESE is the empirical standard deviation of $\hat{\theta}$ across replicates; Coverage is the empirical coverage of nominal 95\% Wald confidence intervals.}
\label{tab:sim_metrics}
\resizebox{\textwidth}{!}{%
\begin{tabular}{@{}llccccccc@{}}
\toprule
\textbf{Scenario} & & \textbf{N} & \textbf{Method} & \textbf{Rel.\ Bias} & \textbf{RMSE} & \textbf{ASE} & \textbf{ESE} & \textbf{Coverage} \\
\midrule
(i)   & Exponential, censored        &  300 & ACSH     & 0.0084 & 0.0160 & 0.0159 & 0.0159 & 0.954 \\
      &                              &  300 & Naive IR & 0.0085 & 0.0158 & 0.0160 & 0.0157 & 0.959 \\
      &                              & 1000 & ACSH     & 0.0004 & 0.0086 & 0.0087 & 0.0086 & 0.960 \\
      &                              & 1000 & Naive IR & 0.0003 & 0.0086 & 0.0087 & 0.0086 & 0.957 \\
\addlinespace
(ii)  & Gamma, no censoring          &  300 & ACSH     & 0.0086 & 0.0106 & 0.0107 & 0.0106 & 0.945 \\
      &                              &  300 & Naive IR & 0.0086 & 0.0106 & 0.0114 & 0.0106 & 0.966 \\
      &                              & 1000 & ACSH     & -0.0005 & 0.0057 & 0.0059 & 0.0057 & 0.958 \\
      &                              & 1000 & Naive IR & -0.0005 & 0.0057 & 0.0062 & 0.0057 & 0.968 \\
\addlinespace
(iii) & Gamma, censored              &  300 & ACSH     & -0.0020 & 0.0128 & 0.0121 & 0.0128 & 0.930 \\
      &                              &  300 & Naive IR & -0.0402 & 0.0130 & 0.0122 & 0.0121 & 0.921 \\
      &                              & 1000 & ACSH     & 0.0011 & 0.0066 & 0.0067 & 0.0066 & 0.951 \\
      &                              & 1000 & Naive IR & -0.0380 & 0.0077 & 0.0067 & 0.0063 & 0.900 \\
\bottomrule
\end{tabular}
}
\end{table}

%------------------------------------
\newpage
%------------------------------------
\newpage
\begin{small}
\begin{table}[ht]
\centering
\caption{Comparison of ACSH (expressed per 100 person-months of event-free follow-up) between DOACs and LMWH over 6 months among patients with cancer-associated VTE in the CANVAS trial. Values are reported as point estimates with corresponding 95\% confidence intervals. Total ACSH denotes the sum of endpoint-specific ACSH estimates across recurrent VTE and major bleeding. \emph{Abbreviations:} ACSH, average cause-specific hazard; DOACs, direct oral anticoagulants; LMWH, low-molecular-weight heparin; VTE, venous thromboembolism.}
\label{tab:acsh_comparison_canvas}
\resizebox{\textwidth}{!}{%
\begin{tabular}{@{}lllll@{}}
\toprule
\textbf{Outcome} & \textbf{DOAC (ACSH, 95\% CI)} & \textbf{LMWH (ACSH, 95\% CI)} & \textbf{Difference (95\% CI)} & \textbf{Ratio (95\% CI)} \\
\midrule
Recurrent VTE
& 1.17 (0.76, 1.82)
& 1.70 (1.16, 2.48)
& -0.52 (-1.35, 0.30)
& 0.692 (0.388, 1.237) \\

Major Bleeding
& 1.06 (0.66, 1.69)
& 1.11 (0.70, 1.76)
& -0.05 (-0.76, 0.66)
& 0.955 (0.495, 1.844) \\

\textbf{Total ACSH}
& \textbf{2.23 (1.48, 2.98)}
& \textbf{2.80 (1.91, 3.70)}
& \textbf{-0.57 (-1.74, 0.60)}
& \textbf{0.796 (0.501, 1.267)} \\
\bottomrule
\end{tabular}
}
\end{table}
\end{small}
%------------------------------------
\subsection*{Funding}
Funding/Support: This work was supported by the National Institutes of Health under award numbers R01GM152499 from the National Institute of General Medical Sciences and R01HL089778 from the National Heart, Lung, and Blood Institute.

This is a secondary analysis of the data from the AFT-28 study (CANVAS), which was funded through a Patient-Centered Outcomes Research Institute (PCORI) Award (CER-1503-29805). All statements in this publication, including its findings, are solely those of the authors and do not necessarily represent the views of the Patient-Centered Outcomes Research Institute (PCORI), its Board of Governors or Methodology Committee. PCORI, the study funder, oversaw research ethics for the CANVAS study, but had no role in the design and conduct of the study; collection, management, analysis, and interpretation of the data; preparation, review, or approval of the manuscript; or decision to submit the manuscript for publication. The Foundation of the Alliance for Clinical Trials in Oncology Foundation, the study sponsor, was responsible for the conduct and management of the study.

\bibliographystyle{biom}
\bibliography{refs}

\clearpage
\appendix
\renewcommand{\thesection}{Appendix~\Alph{section}}
\renewcommand{\thesubsection}{\Alph{section}.\arabic{subsection}}
\setcounter{table}{0}
\setcounter{figure}{0}
\renewcommand{\thetable}{S\arabic{table}}
\renewcommand{\thefigure}{S\arabic{figure}}
\renewcommand{\tablename}{Table}
\renewcommand{\figurename}{Figure}

\section{Asymptotic Results}
\label{app:asymp}
%========================================================
We summarize the regularity conditions and large-sample representations underlying the methods described in the main text (Section 2).

\paragraph{Assumptions} Throughout, we assume that
\begin{enumerate}
\item[(C1)] Independent right-censoring: $C\perp (T,J)$.
\item[(C2)] Positivity on $[0,\tau]$: $G(t)=\Pr(X\ge t)>0$ for all $t\in[0,\tau]$ and $R(\tau)>0$.
\item[(C3)] Each cumulative cause-specific hazard $\Lambda_k(t)$ is finite on $[0,\tau]$.
\item[(C4)] The processes $\{Y_i(t), N_{ki}(t)\}$ satisfy the usual conditions for multiplicative intensity models on $[0,\tau]$, including càdlàg paths and finite variation, ensuring that the Kaplan--Meier and Aalen--Johansen estimators are uniformly consistent and admit asymptotically linear representations on $[0,\tau]$; see, e.g., \cite{Lin1997}.
\end{enumerate}

Under (C1)--(C4), $\hat S(t)$ and $\hat F_k(t)$ are uniformly consistent on $[0,\tau]$, and hence
\[
\hat R(\tau)\xrightarrow{p}R(\tau),
\qquad
\hat\eta_k(\tau)\xrightarrow{p}\eta_k(\tau).
\]

\paragraph{The first order approximation to the CIF Estimator} \label{app:if}
Let
$M_{ki}(t)=N_{ki}(t)-\int_0^t Y_i(u)\,d\Lambda_k(u),$
$M_i(t)=N_i(t)-\int_0^t Y_i(u)\,d\Lambda(u)$
and $G(t)=\Pr(X \ge t).$
It follows from \cite{Lin1997}, we have
\[
\sqrt{n}\{\hat F_k(t)-F_k(t)\}
=
\frac{1}{\sqrt{n}}\sum_{i=1}^n \psi_{F_k,i}(t)+o_p(1),
\]
where
\begin{equation}
\psi_{F_k,i}(t)
=
\int_0^t S(u-)\,\frac{dM_{ki}(u)}{G(u)}
+
\int_0^t \{F_k(u)-F_k(t)\}\,\frac{dM_i(u)}{G(u)}.
\label{eq:psi_Fk_app}
\end{equation}

\paragraph{The first order approximation to the Restricted Mean Event-Free Time Estimator}
For $R(\tau)=\int_0^\tau S(u)\,du,$ and  $\hat R(\tau)=\int_0^\tau \hat S(u)\,du,$
\cite{Zhao2016} showed that 
\[
\sqrt{n}\{\hat R(\tau)-R(\tau)\}
=
\frac{1}{\sqrt{n}}\sum_{i=1}^n \psi_{R,i}(\tau)+o_p(1),
\]
where
\begin{equation}
\psi_{R,i}(\tau)
=
-\int_0^\tau
\left\{R(\tau)-R(u)\right\}
\frac{dM_i(u)}{G(u)}.
\label{eq:psi_R_app}
\end{equation}

\paragraph{The first order approximation to the Log-ACSH Estimator}
Since
\[
\log\hat\eta_k(\tau)=\log\hat F_k(\tau)-\log\hat R(\tau),
\]
delta method yields
\[
\sqrt{n}\{\log\hat\eta_k(\tau)-\log\eta_k(\tau)\}
=
\frac{1}{\sqrt{n}}\sum_{i=1}^n \psi_{k,i}(\tau)+o_p(1),
\]
where
\begin{align*}
\psi_{k,i}(\tau)=&\frac{\psi_{F_k,i}(\tau)}{F_k(\tau)}-\frac{\psi_{R,i}(\tau)}{R(\tau)}\\
=&\int_0^t \frac{S(u)}{F_k(\tau)}\,\frac{dM_{ki}(u)}{G(u)}+\int_0^t \left\{\frac{F_k(u)}{F_k(\tau)}-\frac{R(u)}{R(\tau)}\right\}\,\frac{dM_i(u)}{G(u)}.
\end{align*}
Therefore, 
\[
\sqrt{n}\{\hat{\boldsymbol{\theta}}(\tau)-\boldsymbol{\theta}(\tau)\}
=
\frac{1}{\sqrt{n}}\sum_{i=1}^n \boldsymbol{\psi}_i(\tau)+o_p(1)
\;\to\;
N_q\bigl(\mathbf{0},\boldsymbol{\Sigma}(\tau)\bigr),
\]
in distribution, as $n \to \infty$, where $\boldsymbol{\psi}_i(\tau)=\bigl(\psi_{1,i}(\tau),\ldots,\psi_{m,i}(\tau)\bigr)^\top$ and $\boldsymbol{\Sigma}(\tau)=\mathrm{Var}\{\boldsymbol{\psi}_i(\tau)\}.$ 

\paragraph{Extension to Multiple Non-Terminal Endpoints} For non-terminal event $\ell$, $(X_\ell,K_\ell)$ in Section~2.4 of the main text reduces to a two-cause competing-risk problem with cause 1 being event $\ell$ and cause 2 being death. Applying the preceding approximation results to the log-ACSH for endpoint $\ell$ yields that
$$\sqrt{n}\left\{\log\hat\eta_\ell^{\mathrm{NT}}(\tau)-\log\eta_\ell^{\mathrm{NT}}(\tau) \right\}=\frac{1}{\sqrt{n}}\sum_{i=1}^n \psi_{\ell, i}^{\mathrm{NT}}(\tau)+o_p(1),$$
where 
\[
\psi_{\ell,i}^{\mathrm{NT}}(\tau)
=
\int_0^\tau \frac{S_\ell(u)}{F_{1,\ell}(\tau)}\,\frac{dM_{1, \ell, i}(u)}{G_\ell(u)}
+
\int_0^\tau \left\{\frac{F_{1,\ell}(u)}{F_{1, \ell}(\tau)}- \frac{R_\ell(u)}{R_{\ell}(\tau)}\right\}
\frac{dM_{\ell, i}(u)}{G_\ell(u)},
\]
\[
M_{1, \ell, i}(t)=\mathbb{I}(X_{\ell, i}\le t, K_{\ell, i}=1)-\int_0^{t\wedge X_{\ell, i}} \lambda_{1,\ell}(u)du,
\]
\[
M_{\ell, i}(t)=\mathbb{I}(X_{\ell, i}\le t, K_{\ell, i}>0)-\int_0^{t\wedge X_{\ell, i}}\lambda_\ell(u)du,
\]
\[
\lambda_{1, \ell}(t)=\lim_{\epsilon \downarrow 0} \frac{P(T_\ell\wedge D<t+\epsilon, T_\ell<D \mid T_\ell\wedge D\ge t)}{\epsilon},
\]
\[
\lambda_\ell(t)=\lim_{\epsilon \downarrow 0} \frac{P(T_\ell\wedge D<t+\epsilon\mid T_\ell\wedge D\ge t)}{\epsilon},
\]
$F_{1, \ell}(t)=P(T_\ell<D, T_\ell<t),$
$S_\ell(t)=P(T_\ell\wedge D>t),$
$R_\ell(\tau)=E\{T_\ell\wedge D \wedge \tau\}$
and
$G_\ell(t)=\Pr(X_\ell \ge t).$ 
Joint asymptotic normality across endpoints then follows from the multivariate central limit theorem applied to $\sqrt{n}\{\hat{\boldsymbol{\theta}}^{\mathrm{NT}}(\tau)-\boldsymbol{\theta}^{\mathrm{NT}}(\tau)\}$ with influence function of the $i$th observation being $\boldsymbol{\psi}^{\mathrm{NT}}_i(\tau)=\bigl(\psi_{1,i}(\tau),\ldots,\psi_{\tilde{m},i}(\tau)\bigr)^\top.$

\paragraph{Examples of ACSH interpretation}
\label{app:exp-total-acsh}

To clarify the interpretation of ACSH, consider two independent latent event times corresponding to non-terminal events
\[
T_1 \sim \mathrm{Exp}(\lambda_1), \qquad
T_2 \sim \mathrm{Exp}(\lambda_2),
\]
and a event time corresponding to a terminal event
\[
D \sim \mathrm{Exp}(\lambda_D).
\]
For concreteness, let $T_1$ denote time to recurrent VTE, $T_2$ denote time to major bleeding, and $D$ denote time to death. We assume that $T_1, T_2$ and $D$ are independent.

\paragraph{Terminal Events}
Now consider the standard competing risk setting (Section 2.3). In this setting, we define a single event time as the time to the first event (VTE, bleeding or death), i.e., 
\[
T=\min(T_1,T_2,D),
\]
with the failure cause being the event occurred the first  (VTE, major bleeding first or death). The event-free survival probability is
\[
S(u)=\Pr(T_1>u,\;T_2>u,\;D>u)=e^{-(\lambda_1+\lambda_2+\lambda_D)u}
\]
and
\[
R(\tau)=\int_0^\tau S(u)\,du
=
\frac{1-e^{-(\lambda_1+\lambda_2+\lambda_D)\tau}}{\lambda_1+\lambda_2+\lambda_D}.
\]
The cumulative incidence for VTE as the first event is
\[
F_1(\tau)
=
\Pr(T_1\le \tau,\;T_1<T_2,\;T_1<D)
=
\frac{\lambda_1}{\lambda_1+\lambda_2+\lambda_D}
\Bigl(1-e^{-(\lambda_1+\lambda_2+\lambda_D)\tau}\Bigr).
\]
Thus,
\[
\eta_1(\tau)=\frac{F_1(\tau)}{R(\tau)}=\lambda_1.
\]
Likewise,
\[
\eta_2(\tau)=\lambda_2.
\]
\paragraph{Non-terminal Events}

In this setting, we treat the recurrent VTE and death are two competing risks and the occurrence of major bleeding doesn't affect the ACSH for VTE. In this case, the cumulative incidence of the VTE by $\tau$ is
\[
F_{1,1}(\tau)=\Pr(T_1\le \tau,\; T_1<D)
=
\int_0^\tau \lambda_1 e^{-(\lambda_1+\lambda_D)u}\,du
=
\frac{\lambda_1}{\lambda_1+\lambda_D}\Bigl(1-e^{-(\lambda_1+\lambda_D)\tau}\Bigr).
\]
The corresponding VTE-free survival probability is
\[
S_1(u)=\Pr(T_1>u,\;D>u)=e^{-(\lambda_1+\lambda_D)u},
\]
and the restricted mean VTE-free time is
\[
R_1(\tau)=\int_0^\tau S_1(u)\,du
=
\frac{1-e^{-(\lambda_1+\lambda_D)\tau}}{\lambda_1+\lambda_D}.
\]
Hence,
\[
\eta_1^{\mathrm{NT}}(\tau)=\frac{F_{1,1}(\tau)}{R_1(\tau)}=\lambda_1.
\]
Similarly, the ACSH corresponding to major bleeding $\eta_2^{\mathrm{NT}} (\tau)=\lambda_2.$
The total ACSH is defined as 
\[
\eta^{\mathrm{NT}}_{\mathrm{tot}}(\tau)=\eta_1^{\mathrm{NT}}(\tau)+\eta_2^{\mathrm{NT}}(\tau)=\lambda_1+\lambda_2,
\]
which is simply the sum of two cause-specific hazard rates. 

Thus, we have $$\eta_g(\tau)=\eta_g^{\mathrm{NT}}(\tau), g\in \{0, 1\}.$$
However, despite the equivalence, the scientific interpretations differ. In the standard terminal event setting, $T_1$ competes with $T_2$ and $D$.  In the non-terminal event setting, $T_1$ only competes with $D.$ When the non-terminal events are not mutually exclusive, the interpretation of $\lambda_1+\lambda_2$ as overall disease burden in Section 2.4 is more appropriate.

%=========================================
\newpage
\section{Simulation Results for Cause~2}
\label{app:cause2}

Table~\ref{tab:sim_metrics_cause2} reports the Monte Carlo operating characteristics of the proposed average cause-specific hazard (ACSH) estimator and the naive person-time incidence rate (Naive IR) for Cause~2 under the three data-generating scenarios described in Section~3 of the main text. The simulation configuration is identical to that used for Cause~1 in the main text ($\tau=5$, $N\in\{300,1000\}$, $R=1000$ Monte Carlo replicates per configuration), with true Cause~2 ACSH values of 0.25 in Scenario~(i) and 0.2206 in Scenarios~(ii)--(iii).

The qualitative findings for Cause~2 mirror those for Cause~1 reported in the main text. ACSH is essentially unbiased across all three scenarios with coverage near the nominal 95\% level (0.93--0.95), and ASE closely tracks ESE. The naive IR matches ACSH closely under Scenario~(i) (exponential with censoring) and coincides with it numerically under Scenario~(ii) (Gamma, no censoring), where the two estimators are algebraically identical. Under Scenario~(iii) (Gamma with censoring), the naive IR exhibits a persistent negative relative bias (approximately $-3.0\%$ to $-3.3\%$) that does not attenuate with sample size, and its empirical coverage drops to about 0.90 at $N=1000$, again paralleling the Cause~1 results.

\begin{table}[ht]
\centering
\caption{Monte Carlo operating characteristics of the proposed ACSH estimator and the naive person-time incidence rate for Cause~2 under the three data-generating scenarios described in Section~3 of the main text. Truncation time $\tau=5$, $N\in\{300,1000\}$, $R=1000$ replicates per configuration. True Cause~2 ACSH is 0.25 in Scenario~(i) and 0.2206 in Scenarios~(ii)--(iii). Metric definitions follow Table~2 of the main text.}
\label{tab:sim_metrics_cause2}
\vspace{3mm}
\resizebox{\textwidth}{!}{%
\begin{tabular}{@{}llccccccc@{}}
\toprule
\textbf{Scenario} & & \textbf{N} & \textbf{Method} & \textbf{Rel.\ Bias} & \textbf{RMSE} & \textbf{ASE} & \textbf{ESE} & \textbf{Coverage} \\
\midrule
(i)   & Exponential, censored        &  300 & ACSH     & 0.0084 & 0.0230 & 0.0221 & 0.0229 & 0.933 \\
      &                              &  300 & Naive IR & 0.0081 & 0.0229 & 0.0223 & 0.0228 & 0.938 \\
      &                              & 1000 & ACSH     & 0.0015 & 0.0125 & 0.0122 & 0.0125 & 0.944 \\
      &                              & 1000 & Naive IR & 0.0012 & 0.0124 & 0.0121 & 0.0124 & 0.944 \\
\addlinespace
(ii)  & Gamma, no censoring          &  300 & ACSH     & -0.0007 & 0.0160 & 0.0154 & 0.0160 & 0.948 \\
      &                              &  300 & Naive IR & -0.0007 & 0.0160 & 0.0169 & 0.0160 & 0.961 \\
      &                              & 1000 & ACSH     & -0.0008 & 0.0085 & 0.0085 & 0.0085 & 0.947 \\
      &                              & 1000 & Naive IR & -0.0008 & 0.0085 & 0.0093 & 0.0085 & 0.969 \\
\addlinespace
(iii) & Gamma, censored              &  300 & ACSH     & 0.0051 & 0.0176 & 0.0172 & 0.0176 & 0.950 \\
      &                              &  300 & Naive IR & -0.0298 & 0.0181 & 0.0182 & 0.0169 & 0.939 \\
      &                              & 1000 & ACSH     & 0.0011 & 0.0094 & 0.0095 & 0.0094 & 0.949 \\
      &                              & 1000 & Naive IR & -0.0332 & 0.0116 & 0.0099 & 0.0090 & 0.901 \\
\bottomrule
\end{tabular}
}
\end{table}

%=========================================
\newpage
\section[Robustness of One-Sample ACSH Inference to Hazard Shape, Censoring Intensity, and Event Rarity]{Robustness of One-Sample ACSH Inference to Hazard Shape, Censoring Intensity, and Event Rarity}
\label{app:tier2-robustness}

This appendix extends the main-text simulation study (Section 3.1 and 3.2, Table~2, Table~S1) by varying design factors in the main study: the shape of the latent hazard (increasing versus decreasing), the intensity of independent censoring, and the rarity of the event of interest. 
The data-generating structure is otherwise unchanged---two non-terminal events of interest and a terminal competing event (death) with independent right-censoring, $\tau=5$, $N\in\{300,1,000\}$, and $R=1,000$ Monte Carlo replicates. For each scenario we report the operating characteristics of the proposed ACSH estimator and of the naive person-time incidence rate (Naive IR).

Five scenarios are considered (see Tables~\ref{tab:tier2_robustness} and \ref{tab:tier2_robustness_c2}). Two replace the Gamma latent event times of the main study with Weibull times having an \emph{increasing} hazard (shape $k=2$) or a \emph{decreasing} hazard (shape $k=0.75$); two retain the Gamma latent times but under \emph{light} ($\sim$12\%) or \emph{heavy} ($\sim$66\%) exponential censoring; and one makes the occurence of the first event of interest \emph{rare} with a cause-specific cumulative incidence rate near $3\%$ at $\tau$ (true ACSH $\eta_1\approx0.016$). In all scenarios, time to death always follows $\mathrm{Exp}(0.10)$. 
True ACSH values, which do not depend on the censoring distribution, are obtained by numerical integration and are listed in the caption of Tables~\ref{tab:tier2_robustness} and \ref{tab:tier2_robustness_c2}.

Across all five scenarios the proposed ACSH estimator is essentially unbiased, with the average estimated standard error tracking the empirical standard error and coverage near the nominal $95\%$ level (range $0.927$--$0.965$). The naive incidence rate, by contrast, degrades sharply whenever the latent hazard is non-constant and random censoring is present: its relative bias reaches $-9\%$ under the increasing Weibull hazard and $-15\%$ under heavy censoring, and its $95\%$ Wald coverage falls to only $0.504$ (increasing hazard, $N=1,000$) and  $0.317$ (heavy censoring, $N=1,000$). Under a decreasing hazard the bias of the naive estimator reverses the sign ($+3.7\%$ to $+4.1\%$), and the coverage level of the associated 95\% confidence interval is not satisfactory including the rare event case ($0.883$). These results confirm good operational characteristics of the proposed inference procedure for ACSH, and the failure of the naive incidence rate estimator in the presence of time-varying hazards and random censoring.

One finite-sample caveat is worth noting. Under heavy censoring ($\sim$66\%) and small sample size ($N=300$), the influence function-based standard error estimator was undefined in $3$ of $1000$ replicates for Cause~2 because the risk set was exhausted before $\tau$; these replicates were excluded from the corresponding ACSH average standard error and coverage entries, and the issue does not arise for Cause~1 or  $N=1,000$. This illustrates the practical guidance of Section~5 of the main text that $\tau$ should not exceed the horizon supported by the observed follow-up. The naive incidence rate always returns a finite value but, as shown here, may be severely biased in this regime.

%====
\newpage
\begin{table}[H]
\centering
%\footnotesize
\caption{\textbf{Robustness simulation (Cause~1).} Monte Carlo operating characteristics of the ACSH estimator and the naive person-time incidence rate (Naive IR) for Cause~1 under five data-generating scenarios varying hazard shape, censoring intensity, and event rarity. All scenarios include a terminal competing event (death, $\mathrm{Exp}(0.10)$); $\tau=5$, $N\in\{300,1000\}$, $R=1000$ replicates. True Cause~1 ACSH $\eta_1$: Weibull, increasing hazard ($k=2$) $0.0755$; Weibull, decreasing hazard ($k=0.75$) $0.2044$; Gamma, light censoring ($\sim$12\%) $0.1195$; Gamma, heavy censoring ($\sim$66\%) $0.1195$; Gamma, rare event of interest $0.0160$. Metric definitions follow Table~2 of the main text.}
\label{tab:tier2_robustness}
\vspace{3mm}
\begin{tabular}{@{}rlccccc@{}}
\toprule
\textbf{N} & \textbf{Method} & \textbf{Rel.\ Bias} & \textbf{RMSE} & \textbf{ASE} & \textbf{ESE} & \textbf{Coverage} \\
\midrule
\multicolumn{7}{l}{\textit{Weibull, increasing hazard ($k=2$)}} \\
 300 & ACSH     & -0.0032 & 0.0094 & 0.0091 & 0.0094 & 0.950 \\
 300 & Naive IR & -0.0925 & 0.0110 & 0.0090 & 0.0085 & 0.864 \\
1000 & ACSH     & 0.0005 & 0.0049 & 0.0050 & 0.0049 & 0.965 \\
1000 & Naive IR & -0.0889 & 0.0080 & 0.0049 & 0.0044 & 0.728 \\
\addlinespace
\multicolumn{7}{l}{\textit{Weibull, decreasing hazard ($k=0.75$)}} \\
 300 & ACSH     & 0.0023 & 0.0194 & 0.0192 & 0.0194 & 0.952 \\
 300 & Naive IR & 0.0395 & 0.0214 & 0.0185 & 0.0199 & 0.918 \\
1000 & ACSH     & 0.0020 & 0.0107 & 0.0106 & 0.0107 & 0.948 \\
1000 & Naive IR & 0.0395 & 0.0136 & 0.0101 & 0.0110 & 0.856 \\
\addlinespace
\multicolumn{7}{l}{\textit{Gamma, light censoring ($\sim$12\%)}} \\
 300 & ACSH     & 0.0060 & 0.0109 & 0.0111 & 0.0108 & 0.959 \\
 300 & Naive IR & -0.0058 & 0.0108 & 0.0116 & 0.0107 & 0.967 \\
1000 & ACSH     & 0.0004 & 0.0061 & 0.0061 & 0.0061 & 0.946 \\
1000 & Naive IR & -0.0113 & 0.0062 & 0.0063 & 0.0060 & 0.951 \\
\addlinespace
\multicolumn{7}{l}{\textit{Gamma, heavy censoring ($\sim$66\%)}} \\
 300 & ACSH     & -0.0094 & 0.0188 & 0.0180 & 0.0187 & 0.943 \\
 300 & Naive IR & -0.1534 & 0.0229 & 0.0145 & 0.0137 & 0.712 \\
1000 & ACSH     & 0.0032 & 0.0108 & 0.0102 & 0.0108 & 0.927 \\
1000 & Naive IR & -0.1423 & 0.0188 & 0.0080 & 0.0081 & 0.426 \\
\addlinespace
\multicolumn{7}{l}{\textit{Gamma, rare event of interest}} \\
 300 & ACSH     & -0.0069 & 0.0042 & 0.0042 & 0.0042 & 0.963 \\
 300 & Naive IR & -0.0536 & 0.0040 & 0.0040 & 0.0039 & 0.917 \\
1000 & ACSH     & -0.0045 & 0.0023 & 0.0023 & 0.0023 & 0.962 \\
1000 & Naive IR & -0.0516 & 0.0023 & 0.0022 & 0.0022 & 0.917 \\
\bottomrule
\end{tabular}
\end{table}
%=====
\begin{table}[H]
\centering
\caption{\textbf{Robustness simulation (Cause~2).} Monte Carlo operating characteristics of the ACSH estimator and the naive person-time incidence rate (Naive IR) for Cause~2 under five data-generating scenarios varying hazard shape, censoring intensity, and event rarity. All scenarios include a terminal competing event (death, $\mathrm{Exp}(0.10)$); $\tau=5$, $N\in\{300,1000\}$, $R=1000$ replicates. True Cause~2 ACSH $\eta_2$: Weibull, increasing hazard ($k=2$) $0.1839$; Weibull, decreasing hazard ($k=0.75$) $0.3101$; Gamma, light censoring ($\sim$12\%) $0.2206$; Gamma, heavy censoring ($\sim$66\%) $0.2206$; Gamma, rare event of interest $0.2206$. Under heavy censoring at $N=300$, ACSH entries are evaluated on $997/1000$ replicates (see text). Metric definitions follow Table~2 of the main text.}
\label{tab:tier2_robustness_c2}
\vspace{3mm}
\begin{tabular}{@{}rlccccc@{}}
\toprule
\textbf{N} & \textbf{Method} & \textbf{Rel.\ Bias} & \textbf{RMSE} & \textbf{ASE} & \textbf{ESE} & \textbf{Coverage} \\
\midrule
\multicolumn{7}{l}{\textit{Weibull, increasing hazard ($k=2$)}} \\
 300 & ACSH     & -0.0058 & 0.0136 & 0.0135 & 0.0135 & 0.955 \\
 300 & Naive IR & -0.0927 & 0.0211 & 0.0151 & 0.0125 & 0.826 \\
1000 & ACSH     & 0.0003 & 0.0077 & 0.0075 & 0.0077 & 0.944 \\
1000 & Naive IR & -0.0876 & 0.0176 & 0.0083 & 0.0071 & 0.504 \\
\addlinespace
\multicolumn{7}{l}{\textit{Weibull, decreasing hazard ($k=0.75$)}} \\
 300 & ACSH     & 0.0046 & 0.0272 & 0.0267 & 0.0272 & 0.949 \\
 300 & Naive IR & 0.0412 & 0.0308 & 0.0249 & 0.0280 & 0.903 \\
1000 & ACSH     & 0.0006 & 0.0150 & 0.0147 & 0.0150 & 0.946 \\
1000 & Naive IR & 0.0372 & 0.0192 & 0.0136 & 0.0153 & 0.837 \\
\addlinespace
\multicolumn{7}{l}{\textit{Gamma, light censoring ($\sim$12\%)}} \\
 300 & ACSH     & -0.0010 & 0.0164 & 0.0159 & 0.0164 & 0.943 \\
 300 & Naive IR & -0.0114 & 0.0164 & 0.0173 & 0.0162 & 0.964 \\
1000 & ACSH     & -0.0012 & 0.0086 & 0.0088 & 0.0086 & 0.961 \\
1000 & Naive IR & -0.0117 & 0.0089 & 0.0095 & 0.0085 & 0.958 \\
\addlinespace
\multicolumn{7}{l}{\textit{Gamma, heavy censoring ($\sim$66\%)}} \\
 300 & ACSH     & 0.0128 & 0.0271 & 0.0246 & 0.0269 & 0.928 \\
 300 & Naive IR & -0.1233 & 0.0337 & 0.0214 & 0.0199 & 0.730 \\
1000 & ACSH     & 0.0012 & 0.0141 & 0.0140 & 0.0141 & 0.951 \\
1000 & Naive IR & -0.1283 & 0.0304 & 0.0117 & 0.0110 & 0.317 \\
\addlinespace
\multicolumn{7}{l}{\textit{Gamma, rare event of interest}} \\
 300 & ACSH     & 0.0047 & 0.0179 & 0.0173 & 0.0179 & 0.937 \\
 300 & Naive IR & -0.0306 & 0.0183 & 0.0182 & 0.0171 & 0.934 \\
1000 & ACSH     & 0.0012 & 0.0098 & 0.0095 & 0.0098 & 0.948 \\
1000 & Naive IR & -0.0333 & 0.0120 & 0.0099 & 0.0095 & 0.883 \\
\bottomrule
\end{tabular}
\end{table}

%=========================================
\newpage
\section{Additional Simulation for Two-Sample Contrasts Based on the CANVAS Data}

Because the main motivation for this work is two-sample comparison, Table~\ref{tab:canvas_total_acsh_sim} reports finite-sample performance not only for the Total ACSH summary but also for the endpoint-specific ACSH estimates and their two-sample contrasts (difference and ratio). Simulated samples of size $N\in\{500,1000,3000\}$ were drawn from the observed CANVAS cohort (with replacement, preserving the observed DOAC:LMWH ratio); the largest sample size was included to stress-test the asymptotic approximations underlying the proposed variance estimator. In each simulation replicate we computed arm-specific ACSH for recurrent VTE and major bleeding, the corresponding differences (DOAC $-$ LMWH) and ratios (DOAC / LMWH), and the analogous quantities for Total ACSH. The full-sample ACSH estimates were treated as reference values when evaluating relative bias and coverage. For recurrent VTE these were $0.0117$ (DOAC) and $0.0170$ (LMWH), with difference $-0.0052$ and ratio $0.692$; for major bleeding, $0.0106$ and $0.0111$, with difference $-0.0005$ and ratio $0.955$; and for Total ACSH, $0.0223$ and $0.0280$, with difference $-0.0057$ and ratio $0.796$. Inference for ratio estimands was conducted on the log scale; reported RMSE, ASE, and ESE for ratio rows are therefore on the log scale, and coverage is based on log-scale Wald intervals. All results are based on $R=2000$ simulation replicates.

For the endpoint-specific arm ACSH estimates, relative bias is negligible in absolute terms at all three sample sizes, and empirical coverage is essentially at the nominal 95\% level by $N=3000$ (range $0.944$--$0.953$). The endpoint-specific differences and ratios behave similarly at $N=3000$ (coverage $0.943$--$0.944$); the larger relative-bias magnitudes reported at smaller $N$ for the major bleeding contrasts (up to $-0.11$ on the log-ratio scale and $-0.084$ on the difference scale) simply reflect that the reference difference is very close to zero, so small absolute biases translate into large relative biases.

The Total ACSH summaries are likewise well calibrated. For the Total ACSH arm estimates, the between-arm difference, and the ratio, the average estimated standard error closely tracks the empirical standard error at all three sample sizes, and empirical coverage is near the nominal $0.95$ level. For the Total ACSH ratio the coverages are $0.954$, $0.956$, and $0.945$ at $N=500$, $1000$, and $3000$; for the Total ACSH difference they are $0.956$, $0.956$, and $0.942$; and for the arm-specific Total ACSH they range from $0.935$ to $0.948$. Thus the delta-method variance used to build the Total summary---in which subject-level log-scale influence values for each endpoint are rescaled by the plug-in endpoint ACSH estimates, summed across endpoints, and re-expressed on the required scale---is well calibrated in the sample sizes considered here.

Taken together, Table~\ref{tab:canvas_total_acsh_sim} indicates that the proposed large-sample inference procedure is well calibrated for the endpoint-specific ACSH estimates and their two-sample contrasts across the sample sizes examined, and that the Total ACSH contrasts are likewise well calibrated, with empirical coverage near the nominal level across the sample sizes examined. Because these simulated samples are drawn from the observed CANVAS cohort, the true within-subject dependence between the two endpoints is unknown; the next section therefore complements this analysis with a fully controlled data-generating process in which the cross-endpoint correlation is known by construction, isolating the calibration of the joint endpoint covariance and of the Total ACSH variance.

\newpage
\begin{table}[H]
\centering
\caption{Additional simulation results based on simulated samples drawn from the observed CANVAS cohort ($R=2000$ simulation replicates per $N$). Rows are grouped by endpoint (recurrent VTE, major bleeding, and Total ACSH) and, within each group, report the arm-specific ACSH estimates (DOAC and LMWH), the between-arm difference (DOAC $-$ LMWH, natural scale), and the DOAC / LMWH ratio (inference on the log scale). Reported metrics are relative bias, root mean squared error (RMSE), average estimated standard error (ASE), empirical standard error (ESE), and empirical coverage of nominal 95\% Wald confidence intervals. For ratio rows the reported RMSE, ASE, and ESE are on the log scale.}
\label{tab:canvas_total_acsh_sim}
\vspace{3mm}
\begin{tabular}{llcccccc}
\toprule
Estimand & Contrast & N & Relative Bias & RMSE & ASE & ESE & Coverage \\
\midrule

\multicolumn{8}{l}{\textit{Recurrent VTE}} \\
ACSH        & DOAC        &  500 &  0.0072 & 0.0029 & 0.0029 & 0.0029 & 0.9505 \\
ACSH        & DOAC        & 1000 & -0.0083 & 0.0021 & 0.0021 & 0.0021 & 0.9530 \\
ACSH        & DOAC        & 3000 &  0.0003 & 0.0012 & 0.0012 & 0.0012 & 0.9530 \\
ACSH        & LMWH        &  500 &  0.0059 & 0.0037 & 0.0037 & 0.0037 & 0.9505 \\
ACSH        & LMWH        & 1000 & -0.0020 & 0.0026 & 0.0026 & 0.0026 & 0.9505 \\
ACSH        & LMWH        & 3000 &  0.0012 & 0.0015 & 0.0015 & 0.0015 & 0.9440 \\
Difference & DOAC $-$ LMWH &  500 &  0.0030 & 0.0046 & 0.0047 & 0.0046 & 0.9575 \\
Difference & DOAC $-$ LMWH & 1000 &  0.0123 & 0.0033 & 0.0033 & 0.0033 & 0.9500 \\
Difference & DOAC $-$ LMWH & 3000 &  0.0032 & 0.0020 & 0.0019 & 0.0020 & 0.9425 \\
Ratio      & DOAC / LMWH &  500 &  0.0173 & 0.3368 & 0.3402 & 0.3368 & 0.9635 \\
Ratio      & DOAC / LMWH & 1000 &  0.0302 & 0.2423 & 0.2395 & 0.2421 & 0.9545 \\
Ratio      & DOAC / LMWH & 3000 &  0.0065 & 0.1413 & 0.1369 & 0.1414 & 0.9435 \\
\addlinespace

\multicolumn{8}{l}{\textit{Major Bleeding}} \\
ACSH        & DOAC        &  500 &  0.0076 & 0.0029 & 0.0028 & 0.0029 & 0.9535 \\
ACSH        & DOAC        & 1000 &  0.0021 & 0.0021 & 0.0020 & 0.0021 & 0.9465 \\
ACSH        & DOAC        & 3000 & -0.0036 & 0.0012 & 0.0012 & 0.0012 & 0.9495 \\
ACSH        & LMWH        &  500 &  0.0035 & 0.0030 & 0.0029 & 0.0030 & 0.9530 \\
ACSH        & LMWH        & 1000 &  0.0005 & 0.0021 & 0.0021 & 0.0021 & 0.9505 \\
ACSH        & LMWH        & 3000 &  0.0007 & 0.0012 & 0.0012 & 0.0012 & 0.9465 \\
Difference & DOAC $-$ LMWH &  500 & -0.0839 & 0.0042 & 0.0041 & 0.0042 & 0.9510 \\
Difference & DOAC $-$ LMWH & 1000 & -0.0341 & 0.0030 & 0.0029 & 0.0030 & 0.9475 \\
Difference & DOAC $-$ LMWH & 3000 &  0.0925 & 0.0017 & 0.0017 & 0.0017 & 0.9440 \\
Ratio      & DOAC / LMWH &  500 & -0.1071 & 0.4110 & 0.3896 & 0.4111 & 0.9570 \\
Ratio      & DOAC / LMWH & 1000 & -0.0171 & 0.2810 & 0.2716 & 0.2810 & 0.9480 \\
Ratio      & DOAC / LMWH & 3000 &  0.0918 & 0.1592 & 0.1555 & 0.1592 & 0.9440 \\
\addlinespace

\multicolumn{8}{l}{\textit{Total ACSH}} \\
Total ACSH  & DOAC        &  500 &  0.0074 & 0.0042 & 0.0043 & 0.0042 & 0.9480 \\
Total ACSH  & DOAC        & 1000 & -0.0034 & 0.0031 & 0.0030 & 0.0031 & 0.9350 \\
Total ACSH  & DOAC        & 3000 & -0.0015 & 0.0018 & 0.0018 & 0.0018 & 0.9475 \\
Total ACSH  & LMWH        &  500 &  0.0049 & 0.0052 & 0.0051 & 0.0052 & 0.9460 \\
Total ACSH  & LMWH        & 1000 & -0.0010 & 0.0037 & 0.0036 & 0.0037 & 0.9390 \\
Total ACSH  & LMWH        & 3000 &  0.0010 & 0.0022 & 0.0021 & 0.0022 & 0.9390 \\
Difference & DOAC $-$ LMWH &  500 & -0.0046 & 0.0066 & 0.0067 & 0.0066 & 0.9555 \\
Difference & DOAC $-$ LMWH & 1000 &  0.0083 & 0.0048 & 0.0047 & 0.0048 & 0.9560 \\
Difference & DOAC $-$ LMWH & 3000 &  0.0109 & 0.0028 & 0.0027 & 0.0028 & 0.9415 \\
Ratio      & DOAC / LMWH &  500 & -0.0077 & 0.2677 & 0.2692 & 0.2678 & 0.9535 \\
Ratio      & DOAC / LMWH & 1000 &  0.0153 & 0.1920 & 0.1902 & 0.1921 & 0.9555 \\
Ratio      & DOAC / LMWH & 3000 &  0.0121 & 0.1126 & 0.1094 & 0.1126 & 0.9445 \\
\bottomrule
\end{tabular}
\end{table}

%=========================================
\newpage
\section[Controlled Simulation for the Total ACSH Variance under Known Cross-Endpoint Correlation]{Controlled Simulation for the Total ACSH Variance under Known Cross-Endpoint Correlation}
\label{app:controlled-total}

The two-sample resampling study of the previous section draws from the observed CANVAS cohort, so the true within-subject dependence between endpoints is not known and cannot be varied. This section reports a fully controlled, single-cohort simulation whose purpose is to validate the delta-method variance of the Total ACSH summary of Section~2.4 --- a single-cohort quantity that does \emph{not} reduce to independent two-sample inference. Because the Total ACSH combines the endpoint-specific estimators, its variance depends on the off-diagonal entries of the joint covariance $\boldsymbol\Sigma^{\mathrm{NT}}$ across non-terminal endpoints; a data-generating process in which the cross-endpoint correlation is induced by construction therefore provides a direct check of whether that off-diagonal covariance is correctly estimated. The correlation strength is a controllable design factor.

For each subject we draw a shared gamma frailty $u$ with mean $1$ and variance $v$, and latent non-terminal event times $T_1\mid u\sim\mathrm{Exp}(u\,b_1)$ and $T_2\mid u\sim\mathrm{Exp}(u\,b_2)$ with $b_1=0.10$ and $b_2=0.15$. A shared terminal event (death) time $D\sim\mathrm{Exp}(0.10)$ and an independent censoring time $C\sim\mathrm{Exp}(0.08)$ are also drawn. Each endpoint $\ell\in\{1,2\}$ is then observed as its own two-cause competing-risks process: $X_\ell=\min(T_\ell,D,C)$, with the event of interest recorded when $T_\ell\le\min(D,C)$, death recorded when $D<\min(T_\ell,C)$, and censoring otherwise; occurrences of the other endpoint do not remove a subject from endpoint~$\ell$'s risk set. The shared frailty $u$ and the shared death and censoring times jointly render $\boldsymbol\Sigma^{\mathrm{NT}}_{12}\neq 0$. We consider two correlation regimes, \emph{moderate} ($v=0.5$) and \emph{strong} ($v=1.0$); the induced replicate-level correlation between $\log\hat\eta_1$ and $\log\hat\eta_2$ ranges from about $0.14$ to $0.20$ under moderate frailty and from about $0.17$ to $0.29$ under strong frailty, confirming that the off-diagonal term is a non-trivial validation target. The truncation time is $\tau=5$, sample sizes are $N\in\{500,1000,2000\}$, and $R=1000$ Monte Carlo replicates are used per configuration. Gold-standard ACSH truths, which do not depend on the censoring distribution, were obtained from a single large Monte Carlo draw ($5\times10^6$) from the latent model: under moderate frailty $(\eta_1,\eta_2,\eta_{\mathrm{tot}})=(0.0908,0.1313,0.2220)$, and under strong frailty $(0.0836,0.1174,0.2011)$.

In each replicate we fit the endpoint-specific ACSH estimators on the shared cohort and recover their subject-level influence values, aligned to the original subject order (the within-endpoint time-sort orders differ across endpoints). We then evaluate (a)~the endpoint marginals and (b)~the Total ACSH $\hat\eta_{\mathrm{tot}}=\hat\eta_1+\hat\eta_2$ with the delta-method variance of Section~2.4 and a natural-scale Wald interval. The Total ACSH influence value $\sum_\ell\hat\eta_\ell\,\hat\psi_{\ell,i}$ combines the two endpoints, so its variance incorporates the off-diagonal of $\hat{\boldsymbol\Sigma}^{\mathrm{NT}}$.

Table~\ref{tab:controlled_total_marg} reports the endpoint-specific marginal results and the operating characteristics of the Total ACSH summary. The endpoint-specific estimators were essentially unbiased, with ASE closely tracking ESE and empirical coverage near the nominal $95\%$ level in both correlation regimes (0.935--0.958). The Total ACSH summary showed similar performance. Across $N\in\{500,1000,2000\}$, empirical coverage remained close to $95\%$ in both regimes. Because the Total ACSH variance estimator incorporates the off-diagonal elements of $\hat{\boldsymbol\Sigma}^{\mathrm{NT}}$, these results support the adequacy of the estimated cross-endpoint covariance and the corresponding delta-method inference. These findings are consistent with the CANVAS resampling results in the previous section.

\begin{table}[ht]
\centering
\caption{\textbf{Controlled correlated multi-endpoint simulation.} Monte Carlo operating characteristics of the endpoint-specific ACSH estimators (Endpoint~1, Endpoint~2) and of the Total ACSH summary under two within-subject correlation regimes induced by a shared gamma frailty (moderate, $v=0.5$; strong, $v=1.0$). Data-generating process as described in the text; $\tau=5$, $N\in\{500,1000,2000\}$, $R=1000$ replicates. True values: moderate $(\eta_1,\eta_2,\eta_{\mathrm{tot}})=(0.0908,0.1313,0.2220)$; strong $(0.0836,0.1174,0.2011)$. Metric definitions follow Table~2 of the main text.}
\label{tab:controlled_total_marg}
\vspace{3mm}
\begin{tabular}{@{}llccccccc@{}}
\toprule
\textbf{Frailty} & \textbf{Quantity} & \textbf{N} & \textbf{Rel.\ Bias} & \textbf{RMSE} & \textbf{ASE} & \textbf{ESE} & \textbf{Coverage} \\
\midrule
Moderate & Endpoint 1 &  500 & 0.0031  & 0.0085 & 0.0082 & 0.0085 & 0.940 \\
                          & Endpoint 1 & 1000 & 0.0026  & 0.0059 & 0.0058 & 0.0059 & 0.944 \\
                          & Endpoint 1 & 2000 & -0.0001 & 0.0042 & 0.0041 & 0.0042 & 0.946 \\
\addlinespace
                          & Endpoint 2 &  500 & 0.0045  & 0.0106 & 0.0104 & 0.0106 & 0.935 \\
                          & Endpoint 2 & 1000 & 0.0004  & 0.0075 & 0.0073 & 0.0075 & 0.938 \\
                          & Endpoint 2 & 2000 & 0.0007  & 0.0052 & 0.0052 & 0.0052 & 0.948 \\
\addlinespace
                          & Total ACSH &  500 & 0.0039  & 0.0149 & 0.0142 & 0.0149 & 0.938 \\
                          & Total ACSH & 1000 & 0.0013  & 0.0102 & 0.0100 & 0.0102 & 0.943 \\
                          & Total ACSH & 2000 & 0.0003  & 0.0072 & 0.0071 & 0.0072 & 0.944 \\
\midrule
Strong   & Endpoint 1 &  500 & -0.0003 & 0.0081 & 0.0079 & 0.0081 & 0.941 \\
                          & Endpoint 1 & 1000 & -0.0041 & 0.0055 & 0.0056 & 0.0055 & 0.958 \\
                          & Endpoint 1 & 2000 & -0.0002 & 0.0039 & 0.0039 & 0.0039 & 0.947 \\
\addlinespace
                          & Endpoint 2 &  500 & 0.0062  & 0.0100 & 0.0098 & 0.0100 & 0.941 \\
                          & Endpoint 2 & 1000 & -0.0010 & 0.0068 & 0.0069 & 0.0068 & 0.958 \\
                          & Endpoint 2 & 2000 & 0.0011  & 0.0050 & 0.0049 & 0.0050 & 0.952 \\
\addlinespace
                          & Total ACSH &  500 & 0.0035  & 0.0140 & 0.0140 & 0.0140 & 0.960 \\
                          & Total ACSH & 1000 & -0.0023 & 0.0098 & 0.0099 & 0.0098 & 0.948 \\
                          & Total ACSH & 2000 & 0.0006  & 0.0072 & 0.0070 & 0.0072 & 0.941 \\
\bottomrule
\end{tabular}
\end{table}

\end{document}